\documentclass[preprint,12pt]{aastex}
\usepackage{emulateapj5}
\newcommand{\ee}[1]{\mbox{${} \times 10^{#1}$}}
\newcommand{\eten}[1]{\mbox{$10^{#1}$}}

\newcommand{\degree}{\mbox{$^{\circ}$}}

\newcommand{\kms}{\mbox{km s$^{-1}$}}
\newcommand\cmv{\mbox{cm$^{-3}$}}

\newcommand{\tto}{\mbox{$\rightarrow$}}

\newcommand\smm{submillimeter}

\newcommand{\lsun}{\mbox{L$_\odot$}}
\newcommand{\ta}{\mbox{$T_A^*$}}
\newcommand{\intint}{\mbox{$\int \ta dv$}}

\newcommand{\tk}{\mbox{$T_K$}}
\newcommand{\td}{\mbox{$T_d$}}
\newcommand{\dv}{\mbox{$\Delta v$}}
\newcommand{\vlsr}{\mbox{$v_{LSR}$}}

\newcommand{\mean}[1]{\mbox{$\langle#1\rangle$}} 
\newcommand{\av}{\mbox{$A_V$}} 
\newcommand{\rinf}{\mbox{$r_{inf}$}} 
\newcommand{\rdep}{\mbox{$r_{dep}$}} 
\newcommand{\fdep}{\mbox{$f_{dep}$}} 
\newcommand{\rout}{\mbox{$r_{out}$}} 

\newcommand{\form}{H$_2$CO}
\newcommand{\formi} {H$_2$$^{13}$CO}
\newcommand{\pform} {para-H$_2$CO}
\newcommand{\pformi} {para-H$_2$$^{13}$CO}
\newcommand{\water}{H$_2$O}

\newcommand{\coo}{$^{13}$CO}
\newcommand{\cooo}{C$^{18}$O}
\newcommand{\coooo}{C$^{17}$O}
\newcommand{\hcop}{HCO$^+$}
\newcommand{\hcopi}{H$^{13}$CO$^+$}
\newcommand{\hcopii}{HC$^{18}$O$^+$}
\newcommand{\dcop}{DCO$^+$}
\newcommand{\nthp}{N$_2$H$^+$}
\newcommand{\ntdp}{N$_2$D$^+$}
 
\input{epsf}

\def\plotfiddle#1#2#3#4#5#6#7{\centering \leavevmode
\vbox to#2{\rule{0pt}{#2}}
\includegraphics{#1}}

\newcommand{\etamb}{\mbox {$\eta_{\rm mb}$}}
\newcommand{\thetab}{\mbox{$\theta_b$}}
\newcommand{\jj}[2]{\mbox{$J = #1\rightarrow#2$}}
\newcommand{\jkkjkk}[6]{\mbox{$J_{K_{-1}K_{+1}}
                              = #1_{#2#3}\rightarrow#4_{#5#6}$}}
\newcommand{\sisrf}{\mbox{$s_{ISRF}$}}

\newcommand{\tdr}{\mbox{$T_d(r)$}}
\newcommand{\tkr}{\mbox{$T_K(r)$}}
\newcommand{\nr}{\mbox{$n(r)$}}
\newcommand{\nel}{\mbox{$n_e$}}

\newcommand{\sigmad}{\mbox{$\Sigma_d$}}
\newcommand{\go}{\mbox{$G_0$}}
\newcommand{\chisq}{\mbox{$\chi_r^2$}}

\shorttitle {B335: A Laboratory for Astrochemistry}
\shortauthors{Evans et al.}

\begin{document}

\title{B335: A Laboratory for Astrochemistry in a Collapsing Cloud}
\author{Neal J. Evans II}
\affil{Department of Astronomy, The University of Texas at Austin,
1 University Station C1400, Austin, Texas 78712-0259, U.S.A.}
\email{nje@astro.as.utexas.edu}
\author{Jeong-Eun Lee }
\affil{Department of Astronomy, The University of Texas at Austin,
1 University Station C1400, Austin, Texas 78712-0259, U.S.A.}
\email{jelee@astro.as.utexas.edu}
\author{Jonathan M. C. Rawlings}
\affil{Department of Physics and Astronomy, University College London, Gower
Street, London WC1E 6BT, UK}
\email{jcr@star.ucl.ac.uk}
\author{Minho Choi}
\affil{Taeduk Radio Astronomy Observatory, Korea Astronomy Observatory, 
 Hwaam 61-1, Yuseong, Daejeon 305-348, Korea}
\email{minho@trao.re.kr}

\begin{abstract}

We present observations of 25 transitions of 17 isotopologues of 9 molecules 
toward B335. With a goal of constraining chemical models of collapsing clouds, 
we compare our observations, along with data from the literature, to models
of chemical abundances. 
The observed lines are simulated with a Monte Carlo code, which
uses various physical models of density and velocity as a function of radius.
The dust temperature as a function of radius is calculated self-consistently
by a radiative transfer code. The gas temperature is then calculated at
each radius, including gas-dust collisions, cosmic rays, photoelectric heating,
and molecular cooling. The results provide the input to the Monte Carlo code. 
We consider both {\it ad hoc} step function models for chemical abundances and
abundances taken from a self-consistent modeling of the evolution of
a star-forming core. The step function models can match the observed lines
reasonably well, but they require very unlikely combinations of radial 
variations in chemical abundances. Among the self-consistent chemical models,
the observed lines are matched best by models with somewhat enhanced 
cosmic-ray ionization rates and sulfur abundances. We discuss briefly the steps
needed to close the loop on the modeling of dust and gas, including
off-center spectra of molecular lines.

\end{abstract}


\section{Introduction} \label{intro}

The Bok globule, B335, is a rather round dark globule at a distance of
about 250 pc (Tomita et al. 1979). 
It is perhaps the best case for being a collapsing 
protostar. Observations of CS and H$_2$CO lines (Zhou et al. 1993;
Choi et al. 1995) were reproduced very well with models of
inside-out collapse (Shu 1977). To the extent that such
models may describe the actual density and velocity fields in B335,
this source provides an excellent test bed for astrochemical models. The
only remaining variables in modeling the lines would be the chemical abundances
of the species in question. It is even possible to trace variations in
the abundance as a function of radius because the different parts of
the line profile arise in different locations along the line of sight.
Adding the information from the excitation requirements of different
lines provides a probe of the abundance through the static envelope
and into the collapsing core of the protostar.

On the other hand, the depletion of molecules that is quite apparent in
pre-protostellar cores (e.g., Caselli et al. 2002, Lee et al. 2003)
warns us that molecular lines alone may be misleading.
In the case of B335, Shirley et al. (2002) found that the Shu infall model
that fit the molecular lines (Choi et al. 1995) did not reproduce the
dust emission. They found instead that a power law density model with higher
densities at all radii than the best fit Shu model was needed to fit the
dust emission. We will consider models more similar to the best fitting 
power law as well. 

In general, the molecular lines and dust emission have complementary 
advantages and disadvantages. The lines can be strongly affected by
depletion that varies with radius, while the dust shows no convincing evidence 
so far for variation of opacities with radius (Shirley et al. 2002). 
On the other hand, variation in opacities with radius is also not ruled out,
and the actual value of the opacity at long wavelengths is quite uncertain,
by factors of at least 3 and possibly more. The dust emission is sensitive only
to the column density along a line of sight, while the line emission can in
principle probe the volume density via excitation analysis. Finally, only the
lines can probe the kinematics, but that probe can be confused by depletion
effects (Rawlings \& Yates 2001), and the dust is needed to constrain these effects.
Clearly, the best approach is a unified model for both gas and dust components.

We will present new observations of a large number of species toward B335, using
Haystack Observatory 
and the Caltech Submillimeter Observatory.
We will also present the results of detailed models of radiative transport
in dust to determine the dust temperature for several different physical 
models. Next, we will calculate the gas kinetic temperature, including
gas-dust interactions, cosmic rays, and photoelectric heating.
With these as a basis, we will calculate the 
molecular excitation and radiative transport, using a Monte Carlo code 
(Choi et al. 1995).  A telescope simulation code
will produce model line profiles, given an input model of the
density, temperature, velocity, and abundances as a function of radius, 
for comparison with the observed line profiles.
Based on the comparison, the abundances of various species will be
constrained. We will use step function models for the abundances
and also the results of new calculations of abundances in a cloud collapsing
according to the Shu picture (Lee et al. 2004).

\section{Observations} \label{obs}

We obtained observations of the \hcop\ and \nthp\ \jj10\  lines 
at the Haystack Observatory in 1995 March. Observations
of a large number of lines were obtained at the 
Caltech Submillimeter Observatory in the period 1995 March to 2001 July.
Table \ref{telescopes}  provides the reference frequency for the line, the telescope,
the main beam efficiency (\etamb), the full width at half maximum beam
size (\thetab), the velocity resolution ($\delta v$), 
and the date of observation.
We also provide this information for several observations obtained previously
that are used to constrain the modeling.
The frequencies in Table \ref{telescopes}
 are either those used during observing or those
used later to shift the observed data to an improved rest frequency.
For most lines with hyperfine components, these are the reference frequencies
suitable for a list of hyperfine components that were used to fit lines.
In the case of the \nthp\ \jj10\ line, it is the frequency of the isolated
hyperfine component, best suited for determining the velocity.

In the following sections, we assume that the centroid of B335 is at 
$\alpha = 19^h 34^m 35.4^s; \delta = 07\degree 27\arcmin 24\arcsec$ in
1950 coordinates.  This position 
agrees within 1\arcsec\ with the centroid of the \smm\ emission mapped
with SCUBA (Shirley et al. 2000). This position was originally based on
the position of the millimeter continuum source seen by Chandler \& Sargent 
(1993); more recent interferometric data find a compact component
located 3\farcs6 west and 1\arcsec\ south of this position (Wilner et al. 2000).
At this position, a continuum source is also seen at 3.6 cm, attributed to
a time variable radio jet elongated along the outflow axis (Reipurth et al.
2002).  The difference between our position and the position of the compact
component is not significant for the resolution of these observations.
Some of our data were obtained before we settled on this position.
In cases where we have a map, we may have resampled the data spatially
to synthesize a spectrum at the submillimeter centroid position, 
resulting in a slight degradation of the spatial resolution. 

\section{Results} \label{results}

\begin{figure}
\plotfiddle{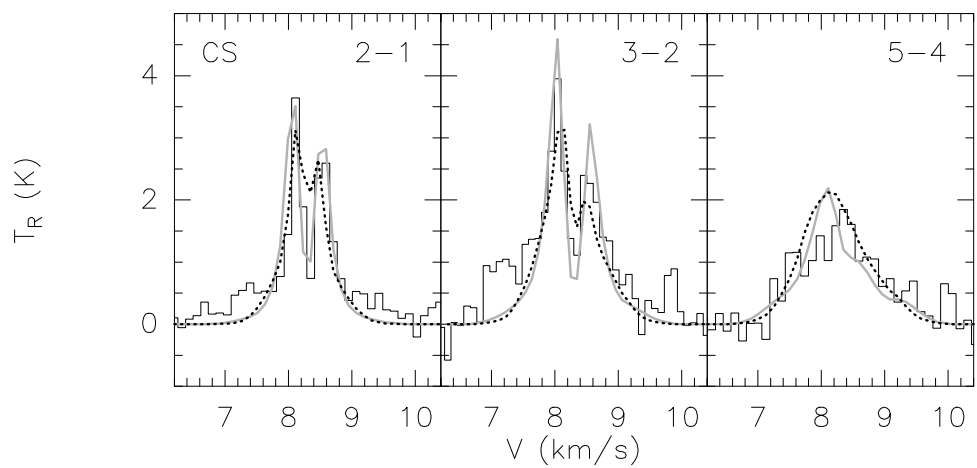}{7.0in}{0}{85}{85}{-250}{-75}
\caption{Observations of the CS lines observed at IRAM (Zhou et al. 1993) 
(black solid histogram). 
The gray solid line is for the model with the step function 
abundance, and the dotted line shows the model with the chemical abundance.
\label{csfig}}
\end{figure}

\begin{figure}
\plotfiddle{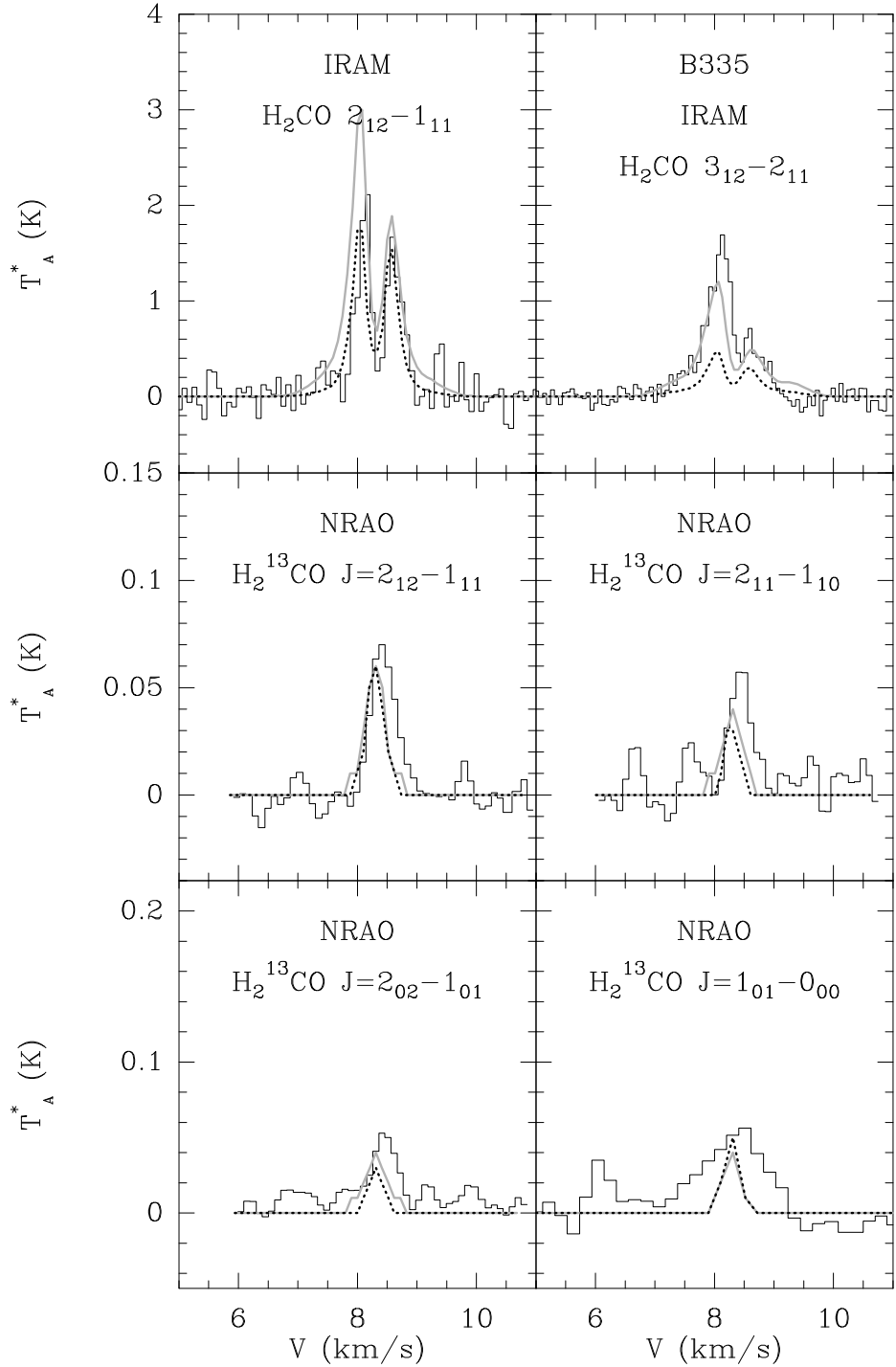}{7.0in}{0}{85}{85}{-250}{-65}
\caption{The observations of \form\ are shown as black solid histograms and 
the models are shown as gray solid lines (step functions) and dotted lines 
(chemical abundances). The telescopes where the data were obtained are 
identified in each panel. The IRAM data were taken from Zhou et al. (1993) 
and the NRAO data were supplied by W. Irvine, based on data in Minh et al. 
(1995). The bottom two panels are \pformi.
\label{form1fig}
}
\end{figure}

\begin{figure}
\plotfiddle{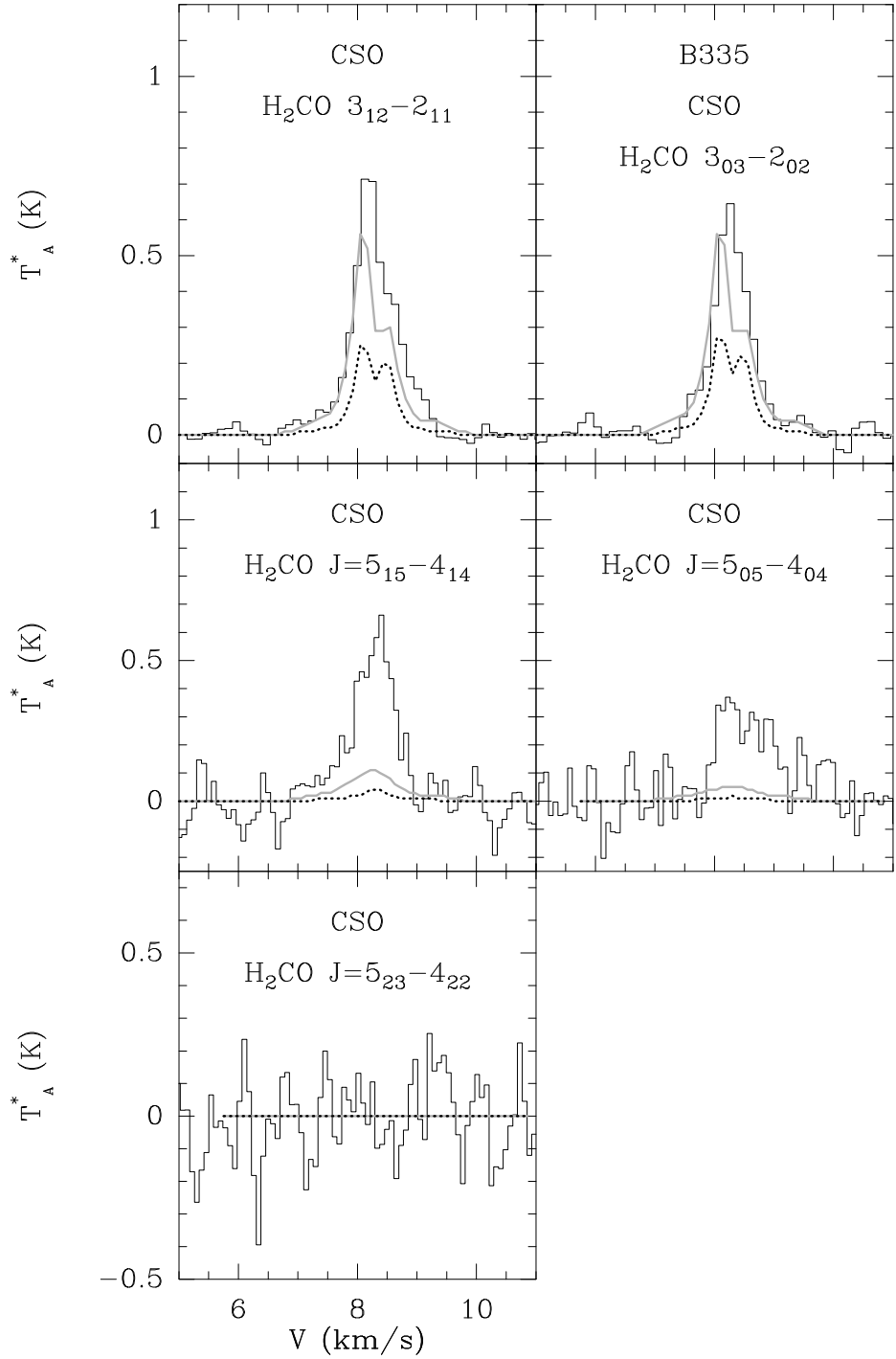}{7.0in}{0}{85}{85}{-250}{-65}
\caption{Further observations of \form\ are shown as black solid histograms 
and the models are shown as gray solid lines and dotted lines. 
The telescopes where the data were obtained are 
identified
in each panel. 
}
\label{form2fig}
\end{figure}

\begin{figure}
\plotfiddle{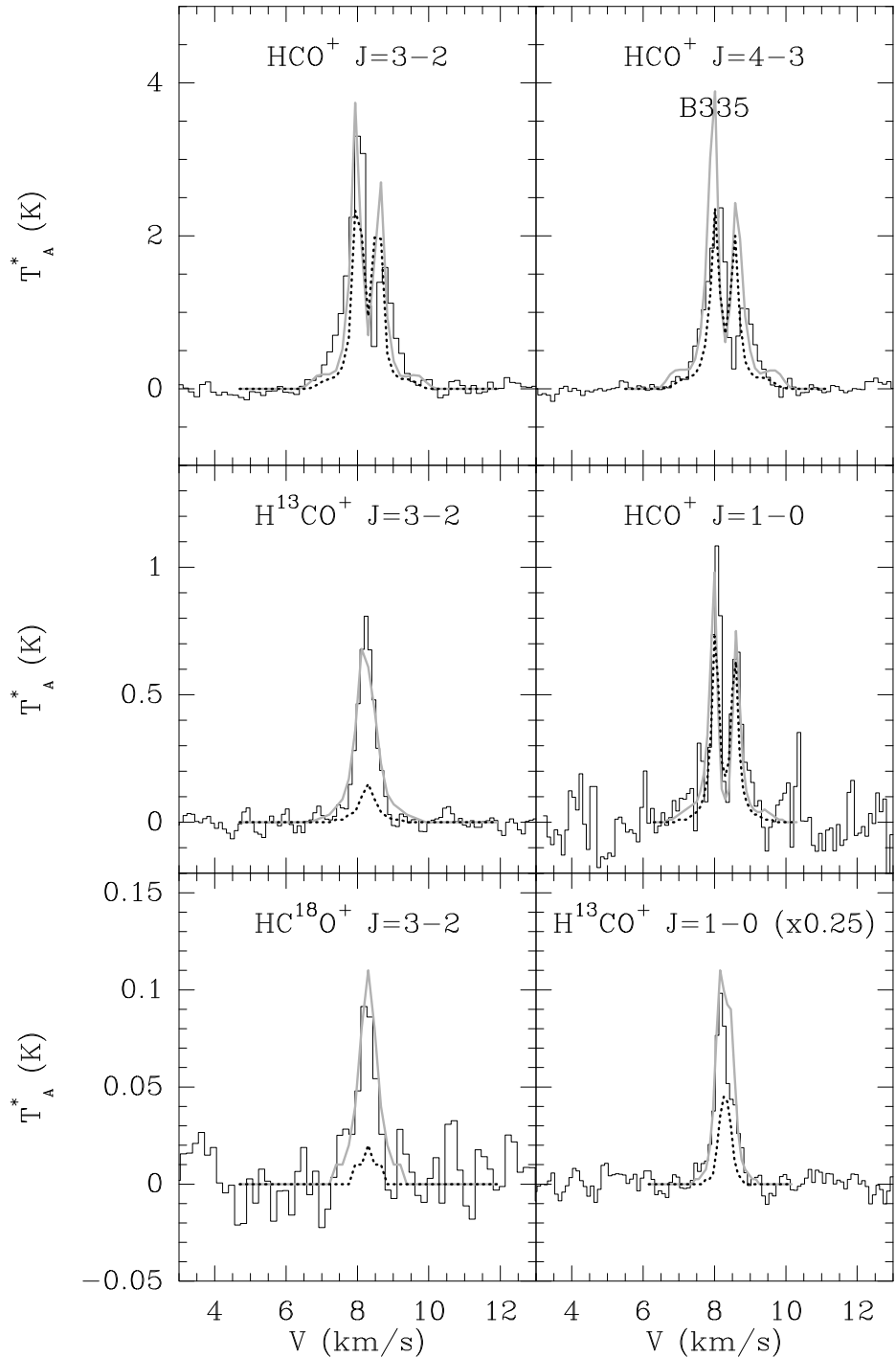}{7.0in}{0}{85}{85}{-250}{-65}
\caption{The observations of \hcop\ are shown as black solid histograms and 
the models are shown as gray solid lines (step functions) and dotted lines 
(chemical abundances). The $J = 1 \rightarrow 0$ lines were obtained at Haystack
and the other lines were obtained at the CSO. Both model and observations of the
\hcopi\ $J = 1 \rightarrow 0$ line have been multiplied by 0.25 to fit them on
the same scale as the HC$^{18}$O$^+$ line.
\label{hcopfig}
}
\end{figure}

\begin{figure}
\plotfiddle{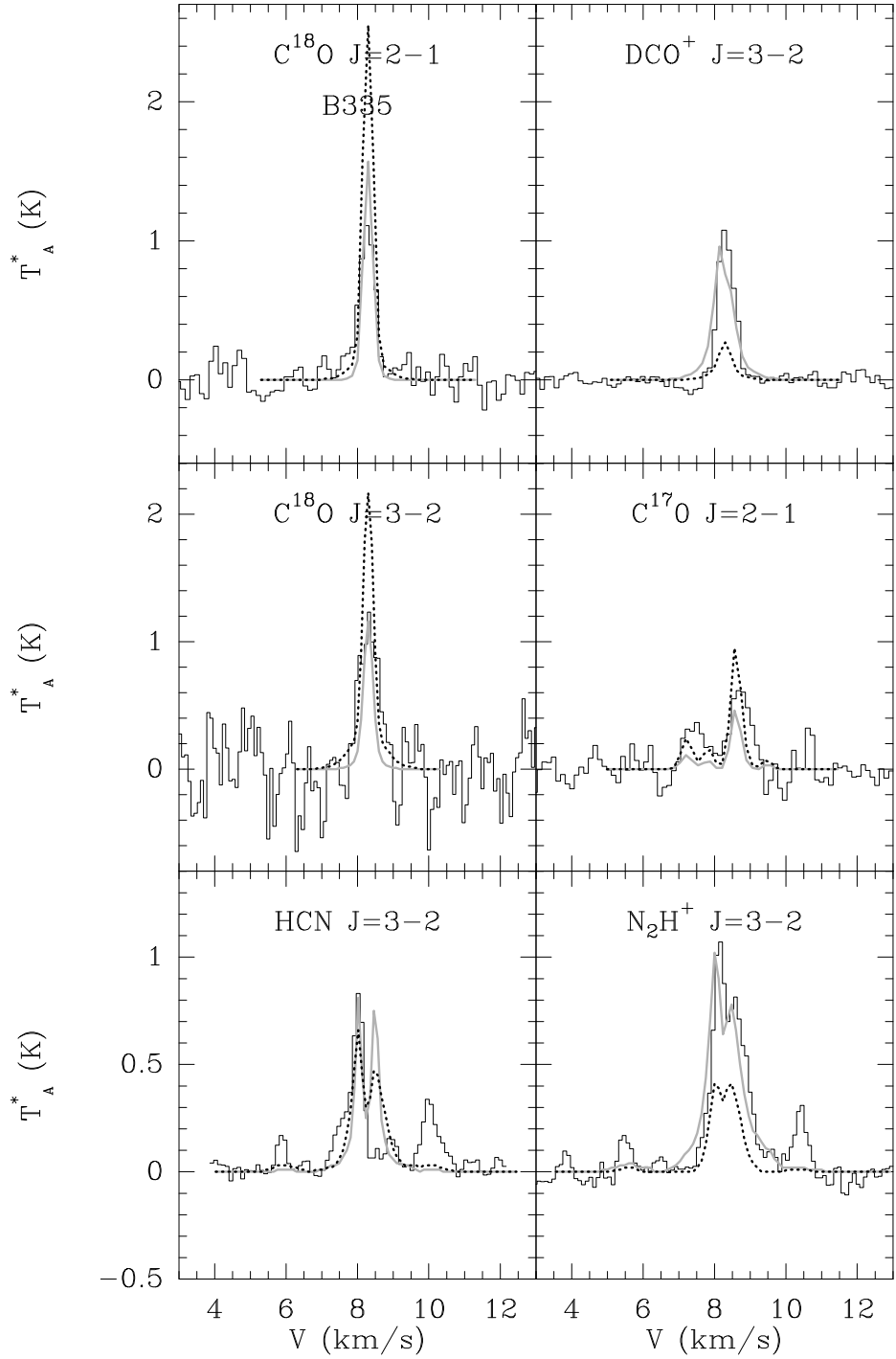}{7.0in}{0}{85}{85}{-250}{-65}
\caption{Selected observations of isotopes of CO, \dcop, HCN,
and \nthp are shown as solid black histograms and the predicted
lines from the model are shown as gray solid lines (step functions) 
dotted lines (chemical abundances). 
\label{cofig}
}
\end{figure}

\begin{figure}
\plotfiddle{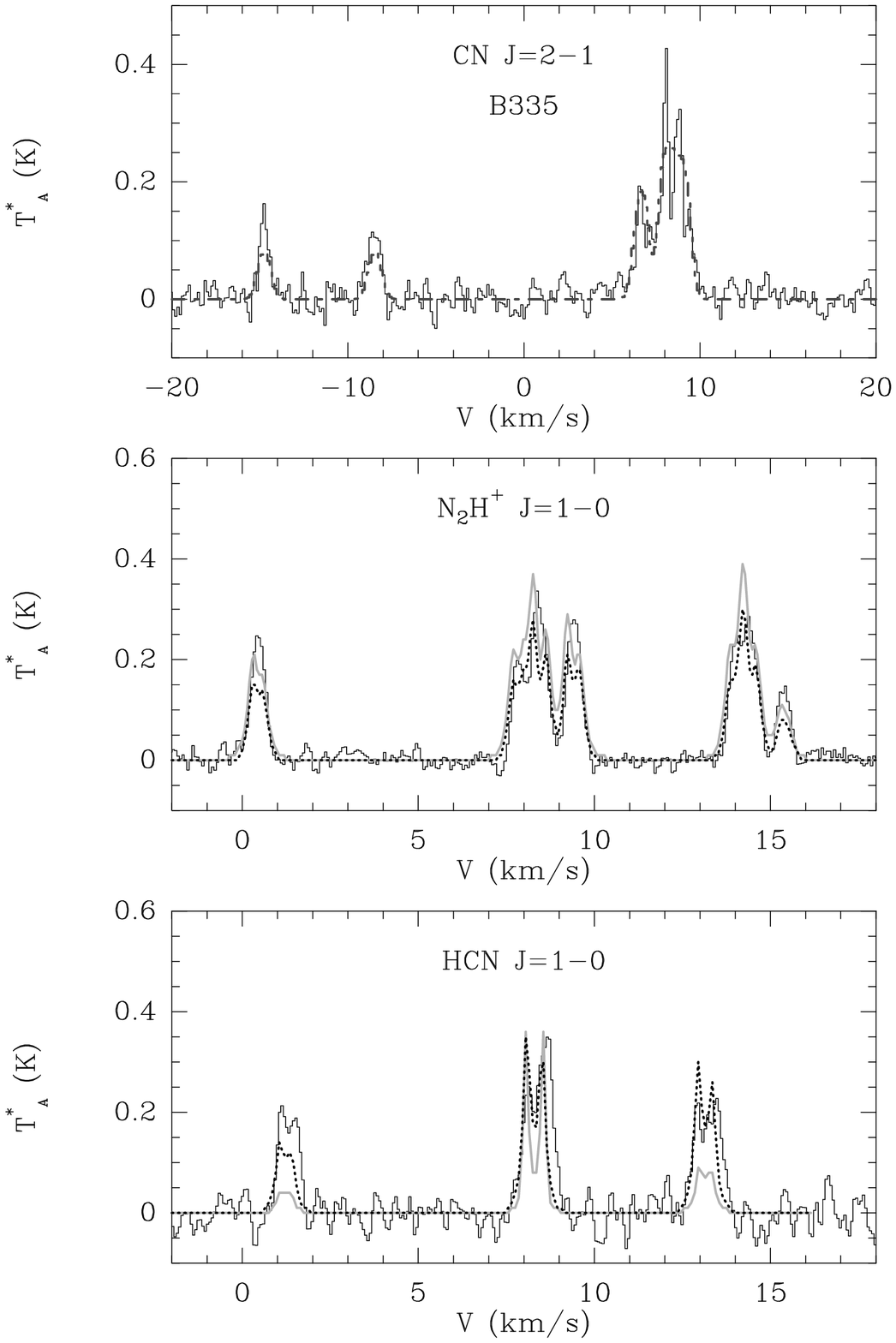}{7.0in}{0}{85}{85}{-250}{-65}
\caption{The observations of the CN \jj21, \nthp\ \jj10\ and HCN \jj10\ lines
are shown as black solid histograms. 
For CN, the dotted line is a fit of Gaussians
to the hyperfine components, not a true model. 
The fit to hyperfine components clearly does not reproduce the shape of the
main group of lines. The dip is centered at 8.35 \kms, indicating that the
main group is self-reversed.
For \nthp\ and HCN \jj10, the gray solid (step functions) and dotted (chemical
abundances) lines are predictions of the radiative transfer model.
\label{cnfig}
}
\end{figure}

\begin{figure}[hbt!]
\plotfiddle{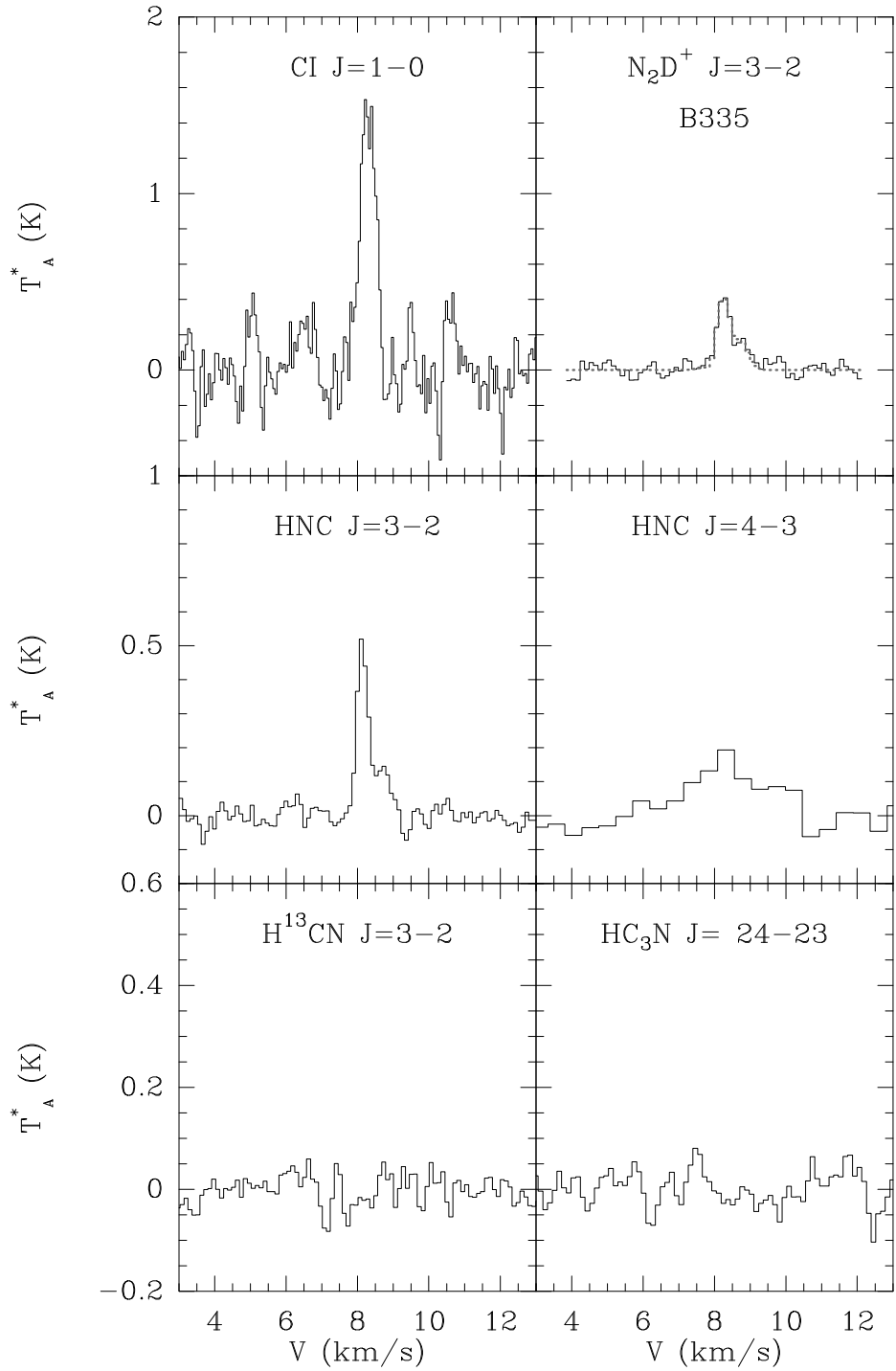}{7.0in}{0}{85}{85}{-250}{-65}
\caption{The observations of lines that are not modeled in
full detail. Some have hyperfine structure. The ones with good
fits using hyperfine components show the fit with a dashed line.
\label{figxx}}
\end{figure}

The primary observational results are presented in Table \ref{resultstab}
and Figure \ref{csfig} to Figure \ref{figxx}. 
The table gives the integrated intensity (\intint), the
peak antenna temperature (\ta), the velocity with respect to the local
standard of rest (\vlsr), and the linewidth (FWHM), \dv. 
For simple, single-peaked lines, these were determined from a Gaussian fit. 
For self-reversed lines without
hyperfine structure (\hcop\ \jj10, \jj32, and \jj43), 
\intint\ is the total area under the full line, \ta\
is the strength of the stronger peak, \vlsr\ is the velocity of the dip,
determined by eye, and \dv\ is \intint\ divided by \ta. For lines with hyperfine
structure (\coooo\ \jj21, \ntdp\ \jj32), 
\intint\ gives the area under all the hyperfine components, \ta\
gives the peak of the strongest, usually blended components, and \vlsr\ and \dv\
come from a fit with all the hyperfine components. For the most complex
situation, lines that are self-reversed, with hyperfine structure, various
strategies were adopted. For CN \jj21, \intint, \ta, and \vlsr\ were determined
as for double peaked lines, but \dv\ was determined from an isolated component.
The spectrum of CN (Figure \ref{cnfig}) clearly shows that the main
hyperfine line is self-reversed.
For \nthp\ \jj10, all line parameters were determined from the isolated component
at the frequency given in Table \ref{telescopes}, as suggested by Lee, Myers,
\& Tafalla (2001).

The observations that are compared to full models 
are shown as solid lines in Figures \ref{csfig} to \ref{cnfig}. 
The CN \jj21\ spectrum in Figure \ref{cnfig} has not been modeled,
and the dashed line is just a fit to the hyperfine components.
Other spectra that are not
modeled in detail are shown in Figure \ref{figxx}. These include spectra with
complex hyperfine splitting that we cannot model in detail and molecules 
without good collision rates.

The HCN \jj32\ line is peculiar in that there is essentially no emission
at velocities that would normally be associated with the red part of the
main hyperfine component. To ensure that this effect was not caused by
emission in the off position (10\arcmin\ west), 
we took a deep integration in the off position.
No emission was seen at a level of 0.08 K.

Single-peaked lines without overlapping hyperfine components provide the 
best measure
of the rest velocity of the cloud. Based on those lines least likely to
be optically thick, the cloud velocity is $\mean{\vlsr} = 8.30\pm0.05$ \kms. 
For self-reversed lines, the mean velocity of the dip (determined by eye) is 
$\mean{v_{dip}} = 8.41\pm0.06$ \kms. 
All the values for $v_{dip}$ exceed those for \mean{\vlsr}, by amounts
ranging from 0.05 \kms\ to 0.25 \kms. The mean shift,
$\mean{v_{dip}-\mean{\vlsr}} = 0.11\pm0.07$. Also, lines
from higher $J$ levels have higher $v_{dip}$ than those from lower $J$ levels, 
suggesting that the dip arises partially from inflowing gas.
The three lines of \hcop, for example, have their dip at increasing velocity,
with the \jj43 showing $v_{dip} = 8.55$ \kms. This progression is similar 
to a pattern seen in CS lines toward IRAM04191 by Belloche et al. (2002).

\section{The Modeling Procedure}\label{modeling}

We use the extensive observations described above to test models of
the source. All the models are spherical models with smooth (non-clumpy)
density distributions. We focus on inside-out collapse models, though
we discuss some variations on this basic model. All models include 
self-consistent calculations of the dust and gas temperature distributions
(\S \ref{temps}) and calculations of the molecular populations, radiative
transport, and line formation (\S \ref{mc}). Two kinds of models of the
abundances as a function of radius are used: step function models, and
abundances from an evolutionary chemical calculation (Lee et al. 2004),
as described in \S \ref{chem}.

\subsection{Determining Temperatures} \label{temps}

The first step in comparing a physical model to observations is to 
determine the temperatures that correspond to a particular density
distribution. The dust temperatures can be calculated self-consistently
for a particular density distribution by various radiative transfer codes.
We used the code of Egan et al. (1988) and the techniques
described by Shirley et al. (2002) for constraining parameters. 

We assumed that dust opacities are given by column 5 of the table in
Ossenkopf \& Henning (1994), known as OH5 opacities, because these have
been shown to match many observations of star forming cores (e.g.,
Shirley et al. 2002).  One difference
between the models by Shirley et al. and the current work is in the 
treatment of the interstellar radiation field (ISRF). In the previous work,
we decreased the strength of the ISRF by a constant factor (\sisrf) at all 
wavelengths (except for the contribution of the cosmic microwave background). 
For B335, we used $\sisrf = 0.3$. In the present work, we instead attenuate
the ISRF using the Draine \& Lee (1984) extinction law and assuming 
$\av = 1.3$ mag. This procedure affects short wavelengths much more than long
wavelengths, leading to a somewhat less pronounced rise in dust temperature
toward the outside of the cloud. The choice of $\av = 1.3$ mag is somewhat
arbitrary, but it accounts for the fact that molecules require some 
dust shielding. It will also produce consistent results when we consider
the gas energetics.

Shirley et al. assumed an outer radius of 60,000 AU for most B335 models.
We will mostly use an outer radius of 0.15 pc (31,000 AU), as used by Choi
et al. (1995). Studies of the extinction as a function of impact parameter
from HST/NICMOS data are consistent with an outer radius of about this
size (Harvey et al. 2001), in the sense that the extinction decrease with
radius blends into the noise at that radius. The choice of outer radius
makes little difference in most models. The inner radius for the dust models
is taken to be \eten{-3} of the outer radius for the dust models in order to
capture the conversion of short-wavelength radiation from the forming star
and disk to longer wavelength
radiation. The stellar temperature is set to 6000 K, but this choice 
is completely
irrelevant because of the rapid conversion to longer wavelength radiation.
The luminosity was set to 4.5 \lsun, which provided the best fit to a Shu
model in Shirley et al. (2002). 

Once we have a dust temperature distribution, \tdr, and a density distribution,
\nr, we can compute the gas temperature distribution, \tkr. This was done 
with a gas-dust energetics code written by S. Doty [see Doty \& Neufeld (1997)
and the appendix in Young et al. (2004) for descriptions]. 
This code includes energy transfer
between gas and dust, heating by cosmic rays and the photoelectric effect
with PAHs, and molecular cooling. 

The gas-dust energy transfer via collisions depends on the total grain
cross section per baryon, averaged over the distribution of grain sizes
(\sigmad). Following the discussion in the Appendix of Young et al. (2004),
we take this value to be 6.09\ee{-22} cm$^2$. The cosmic ray heating depends
on the cosmic ray ionization rate ($\zeta$), which we take to be
3\ee{-17} s$^{-1}$ (van der Tak \& van Dishoeck 2000). 
The photoelectric heating follows the
equation of Bakes \& Tielens (1994), which includes heating from the 
photoelectric effect on very small grains. The rate depends on the strength
of the ultraviolet portion of the ISRF and the electron density.
Because this heating is only important on the outside of the cloud, we set
the electron density to 1\ee{-3} \cmv. The radiation field is assumed to
be attenuated by the surrounding medium according to $\tau_{UV} = 1.8 \av$; 
with $\av = 1.3$, the scale factor
for the ISRF impinging on our model's outer radius is $\go = 0.1$. Once inside
the cloud, the radiation is attenuated according to a fit to the attenuation
produced by the dust assumed to be in the cloud. 

\begin{figure}[hbt!]
\plotfiddle{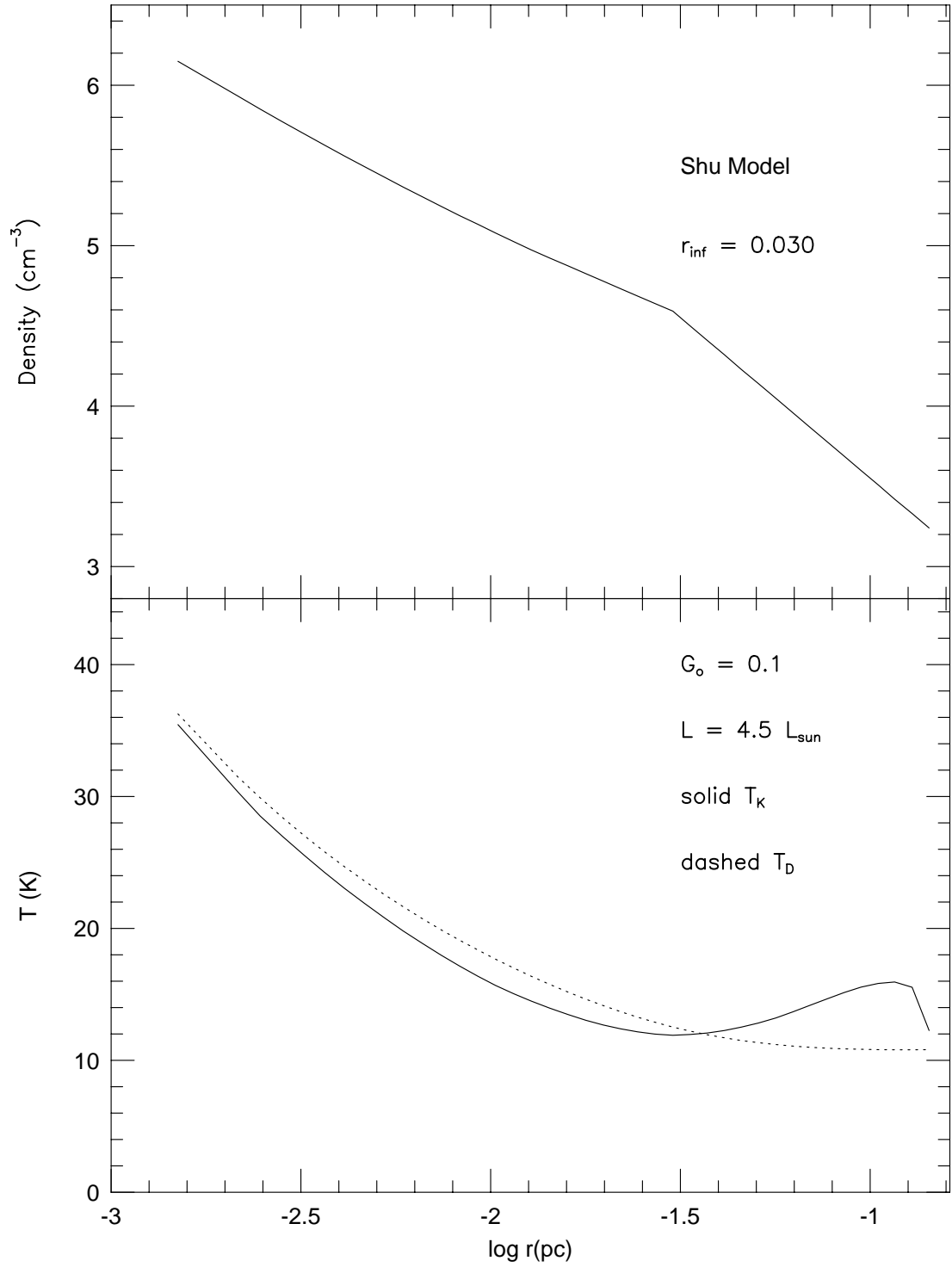}{7.0in}{0}{85}{85}{-250}{-65}
\caption{The density and temperatures for gas and dust plotted as a function
of radius for the standard model.
\label{dentemp}}
\end{figure}

The result of these calculations is shown for a typical model in Fig.
\ref{dentemp}. For small radii, $\tk \approx \td$, as is usually assumed, but
\tk\ falls below \td\ with increasing radius, as the density becomes too low
for collisions with dust to maintain the kinetic temperature at the dust
temperature. Then, at some radius, \tk\ rises as photoelectric
heating takes over, and $\tk > \td$. The downturn in \tk\ at the cloud edge
appears to be real and caused by the cooling lines of CO becoming optically
thin (see Young et al. 2004). 
However, this drop in temperature has no appreciable 
effect on the resulting line profiles.

The amount of photoelectric heating is the least certain of these inputs,
as the external attenuation has a large effect on how warm the outer cloud
gets. To constrain \go, we modeled the lower three lines of CO and compared
to data in the literature (e.g., Goldsmith et al. 1984, 
Langer, Frerking, \& Wilson 1986). 
To avoid producing a CO \jj10\ line that exceeded
the observations, \go\ definitely needed to be decreased from unity. The value
of $\go = 0.1$ provided the best match, and this was actually used to 
constrain the external extinction to $\av = 1.3$. Changes by a factor of
2 in \go\ (or $\pm 0.4$ in \av) produced CO lines that differed from
the observations by about 30\%, while having no appreciable effect on
the lines of other species.  While other variables are uncertain 
in the photoelectric heating, the attenuation of the ultraviolet radiation
from the ISRF is the most important variable; comparison to observations of
CO readily constrain it. The results are reasonable; one does not expect
significant molecular gas for $\av < 1$ (van Dishoeck \& Black 1988).
One could trade off the value of \sisrf\ and the external extinction,
as long as the effective \go\ is not too different from 0.1. Constraining
these separately is difficult (Shirley et al. 2004) and not particularly
relevant for this paper.

The cooling
rates (Doty \& Neufeld 1997) depend primarily on the CO abundance and the
width of the lines (through trapping); we assume $X(CO) = 7.4\ee{-5}$ and
$b = 0.12$ \kms, except for some tests described below.
We note in passing the dangers of simplistic interpretation of observed
CO lines; turning the observations into a kinetic temperature would lead
one to conclude that \tk\ is constant within the cloud, while it clearly
is not.

The parameters that describe the standard physical model are summarized in 
Table \ref{physmod}.

\subsection{Chemical Modeling} \label{chem}

Two kinds of abundance models are employed. The first is strictly
{\it ad hoc}, using a step function to describe the abundance of each
species as a function of radius. These models have three free parameters
per species: $X$, \rdep, and \fdep.
The abundance in the outer parts of the cloud ($X$) is assumed to
decrease inside a depletion radius (\rdep) by a factor (\fdep). 

The second are true chemical models, based on the calculations 
presented by Lee et al. (2004). These calculations follow the chemical evolution
through an evolutionary sequence that includes at each step a self-consistent
calculation of the dust and gas temperatures, using the techniques
described in the previous section. The evolutionary model assumes 
a slow build-up in central density, via a sequence of Bonnor-Ebert spheres,
to the point of a singular isothermal sphere, at which point, an inside-out
collapse (Shu 1977) is initiated. After this point, the chemistry is 
calculated for each of 512 gas parcels as it falls into the central region.
Thus, gas inside the infall radius carries some memory of the conditions
from farther out. 
We adopt the model of Young and Evans (2004) for the evolution of luminosity
in order to calculate the evolution of dust temperature. 
Physical parameters in the model are selected to have a total internal 
luminosity and a dust temperature profile similar to those obtained from the 
dust modeling of B335, at the time step of $r_{inf}=0.03$ pc.
The model core is assumed to stay in the same environment through its evolution
with the same $A_V$ and $G_0$ calculated in the previous section.
The chemical calculation includes the interaction between gas and dust grains
as well as gas-phase reactions, but the surface chemistry is not considered 
in the calculation.
For details of the chemical evolution model, refer to Lee et al.  (2004). 

For both types of models, isotope ratios were constrained so that the
abundance was only a free parameter for the whole of the isotope complex.
\dcop\ was the exception, as it is subject to large fractionation effects.
Assumed isotope ratios are the same as those used by J{\o}rgensen et al.
(2004) and are given in Table \ref{isotopes}. Wouterloot et al. (2004)
have recently suggested a slightly higher ratio for $^{18}$O/$^{17}$O of
4.1, but this value would fit our data on the CO isotopologues somewhat worse.

\subsection{Modeling of Line Profiles} \label{mc}

The line profiles were modeled with a Monte Carlo code (mc) to calculate 
the excitation of the energy levels and a virtual telescope program (vt) to
integrate along the line of sight, convolve with a beam, and match the
velocity resolution, spatial resolution, and main beam efficiency of 
the observations (Choi et al. 1995). All lines were assumed to be centered
at 8.30 \kms, based on the average of optically thin lines.

The input physical conditions (density, 
temperature, and velocity fields) were taken from the physical model
being tested, using the results of the gas-dust energetics code for \tkr.
Models require input data about each molecule, as well as about the source.
For CS, we used collision rates from Turner et al. (1992).
For \hcop\ and \nthp, we used collision rates supplied by B. Turner, based on 
his extension of previously calculated rates to higher temperatures and energy
levels (Turner 1995). For \form\ and \pform, we used rates computed by 
Green (1991). Rates for HCN came from Green \& Thaddeus (1974) and those for
CO from Flower \& Launay (1985).
In some cases, rates have been extrapolated to lower temperatures.

For \coooo, HCN, and \nthp, the lines have hyperfine structure that
is partially resolved. For these lines, mc and vt models were run separately
for each clearly resolved hyperfine component, with abundances adjusted to
simulate the fraction of the transition probability in that component; the
results were added to make the final simulated line. Components separated
by less than the $1/e$ width of the velocity dispersion 
were aggregated into a single component; the aggregated components
are listed in Table \ref{hpftab}. This
procedure captures the essence of the hyperfine splitting, but it is not
rigorous because trapping is not handled correctly when there is
 partial line overlap (see Keto et al. 2004).

All models were run with 40 shells. The inner radius was 2\ee{-3} pc,
corresponding to 1\farcs7 at a distance of 250 pc. This radius is larger than
the inner radius for the dust models because the molecular lines
are not sensitive to emission from very small scales because of beam dilution.
The convergence criterion for populations was set to 10\% for finding the 
region of best-fitting parameters. 
Final models were run with a 2\% convergence criterion to ensure
accuracy; differences between these models and those run with 10\% accuracy
were small.  The minimum fractional population tested for
convergence was $10^{-6}$. For an explanation of these criteria,
see the Appendix in Choi et al. (1995). 

\section{Inside-out Collapse Models}

The physical properties of the standard model are given in Table \ref{physmod}.
The standard physical model is the inside-out collapse model (Shu 1977)
that best matched (Choi et al. 1995) the CS and \form\ data taken by Zhou et 
al. (1993). Choi et al. (1995) modeled CS and \form\ lines from the
IRAM telescope, assuming constant abundances,
and found a best fit $\rinf = 0.03$ pc. This was a compromise,
as \form\ favored smaller \rinf\ than did CS.

While the CS data were still well matched with constant abundance,
the new data on more \form\ lines suggested enhanced abundances
on small scales, as did the \hcop\ data.
As a result, we tested step function abundance models.

\subsection{Step Function Abundances}\label{step}

To avoid too many free parameters, we required that $\rdep = \rinf$.
While this particular choice has no theoretical justification, it 
leaves only two free parameters per species. With the constraints on
isotope ratios in Table \ref{isotopes}, we are left with 15 free parameters
for 8 species, including the special case of \dcop, explained below.
The abundances in Table \ref{abun} are those that fit
the current data reasonably well, as judged by eye and statistical 
measures. We calculated both the reduced chi-squared (\chisq) and 
the absolute deviation  
($ AD = \sum_i|T_A^*(model;i)-T_A^*(obs;i)|/N$)
over the line profiles. The absolute deviation is more influenced
by strong lines for which the shape is important to match, 
so we use it primarily, though
the \chisq\ criterion does not differ in the choice of best model.
We have not run a complete grid of models; instead, we employed
some judgment to locate regions of parameter space with decent fits
to the line profiles. Once reasonably good fits
were obtained, both $X$ and \fdep\ were varied by factors of 3 in 
each direction, showing substantially worse fits. These parameters
should be considered constrained at that level.
The abundances in Table \ref{abun} are those for the 
best-fitting step functions, and
the predicted line profiles are shown in Figures \ref{csfig} to \ref{cnfig} 
as gray lines. The values of $AD$ in Table \ref{abun} are an average over
all the lines for all isotopologues of that species.

The CS lines were still matched best with constant abundance. 
On the small scales (0.003 pc) probed by interferometers, the CS is clearly
depleted in the envelope and the \jj54\ emission arises from a clump that
is offset from the central source (Wilner et al. 2000). A model with
CS depleted by a factor of 10 for $\rdep = 0.003$ pc showed no appreciable
effect on the \jj21\ or \jj32\ lines, but it did decrease the predicted
\jj54\ intensity slightly.

The \form\ abundance that best fits the data is slightly higher 
than was found by Choi et al. (1995), mostly to improve the fit to the
lines of \formi\ and \pformi. 
We also increased the abundance of both \form\ and \pform\ in
the inner parts of the cloud to improve the fit to our new CSO observations of
the higher-J lines, whereas Choi et al. (1995) had found a constant 
abundance to be satisfactory. Even so, we do not reproduce the very high
excitation \form\ lines, indicating that a warm, dense region must exist that
is not predicted by the basic model.

The abundance of \form\ listed in Table \ref{abun}
is actually the abundance of ortho-\form.
Minh et al. (1995) found that ortho-\form /\pform\ was
1.7 in B335, assuming a uniform cloud. 
Our modeling, which employs density, temperature, and velocity
gradients, confirms that this ratio works well in reproducing the observations,
but we have not determined the range of acceptable values.
Minh et al. (1995) noted that this ratio was consistent with ortho-para 
equilibration on cold dust grains and suggested that the 
gas-phase \form\ in B335 had formerly
resided on dust grains. In this picture, they suggested that warming by the
newly-formed star or by shocks had liberated the \form\ from the dust grains.
Our model for the dust temperature indicates that \td\ stays below 20 K until
$r < 0.006$ pc (6\arcsec), where any \form\ would be beam diluted. Thus, other
means for releasing the \form\ from dust mantles should be explored.

The lines of \hcop\ and its isotopologues are best matched with a model
with increased abundances inside \rinf. \hcop\ is clearly quite abundant
in B335, as witnessed by the detection of \hcopii.
The observed \jj32\ and \jj43\ lines of \hcop\ are
somewhat weaker than the models predict and the dip is shifted to the
red (\S \ref{variations}). 
The abundance of \dcop\ was treated as a free parameter but \rdep\ and
\fdep\ were constrained to the same value as for \hcop; the best fit
was obtained for \hcop/\dcop\ of 55.

The \cooo\ lines were fitted best with decreased abundances (but
only by a factor of 3) inside \rinf. With the enforced \cooo/\coooo\
ratio of 3.5, the \coooo\ model line is a bit weaker than the observed
line, but the data are rather noisy. Using the standard isotope ratio,
the best fit abundance of \cooo\ would imply $X(CO) = 4\ee{-5}$. This
abundance is substantially less than expected from chemical models and
even less than what we assume in our calculation of cooling rates
(\S \ref{temps}). To see the consequences, we ran a model with the
abundance adjusted to this value in the calculation of cooling rates.
The value of \tk\ in the outer parts of the cloud was increased by a few
degrees, but the effect on most molecular lines was very small, indicating
that the best fit is not affected by this slight inconsistency. 
The CO line predictions were exceptions, as these lines actually got 
{\it stronger} with {\it decreased} CO abundance because of the higher \tk\ in 
the relevant layers of the cloud.

The two lines of \nthp\ both have hyperfine structure and our method
for dealing with this is only approximate. Nonetheless, the \jj10\ transition
is matched reasonably well (Fig. \ref{cnfig}) with a factor of 10 increase
in abundance inside \rinf. In contrast, the satellite
hyperfine lines of the \jj32\ line are clearly stronger than the models
can explain.

The most troublesome species was HCN. The satellite hyperfine lines
of both \jj10\ and \jj32\ transitions are much stronger than the models
can account for, even with a very large HCN abundance. Still larger abundances
predicted lines of the stronger hyperfine components that were much stronger
than observed. In addition, the HCN \jj32\ line is very peculiar, with the
red side of the line essentially missing, indicative of a deep absorption 
layer. To try to match some of these features within the constraints of our 
model, we depleted HCN by a factor of 10 inside \rinf. This helped, but the fits
are still poor. The fact that the H$^{13}$CN \jj32\ line was not detected
makes the strength of the hyperfine satellite lines (Fig. \ref{figxx})
even harder to understand.

\subsection{Variations in the Physical Model}\label{variations}

\begin{figure}
\plotfiddle{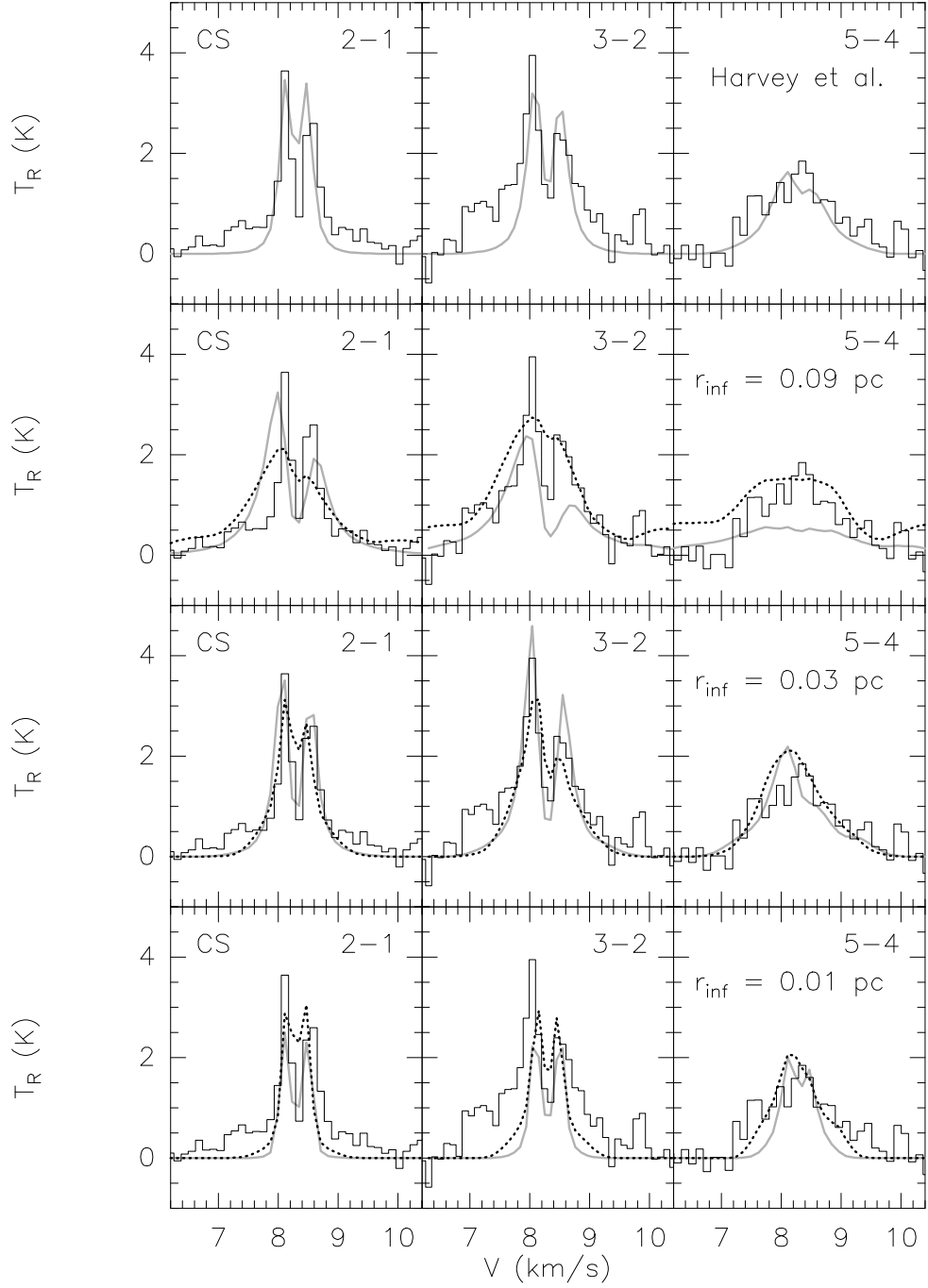}{7.0in}{0}{85}{85}{-250}{-65}
\caption{The observed CS lines from Zhou et al. (1993) are plotted with
the line profiles predicted by various models. The step function models
are gray and the evolutionary models are dotted. The bottom panel shows the
best model with $\rinf = 0.01$ pc; the second panel from the bottom shows
the best model with $\rinf = 0.03$ pc (same as Fig. \ref{csfig}); the third
panel from the bottom shows the best model with $\rinf = 0.09$ pc; and
the top panel shows the best model with the $\rinf = 0.03$ pc, but with
densities enhanced by a factor of 5 everywhere, as favored by Harvey
et al. (2001).
\label{cscomp}
}
\end{figure}

With the additional freedom of the step function abundance profile,
is $\rinf = 0.03$ pc still the best model? This question was explored to
a limited degree; for each new \rinf, the abundances of each species
were optimized, but the shape of the step function was not allowed to
change, except for CS, where changes in \fdep\ were allowed. 
For modest changes ($\rinf = 0.02-0.04$ pc), the overall fits were
not much worse. As found by Choi et al. (1995), 
the CS favored $\rinf = 0.03$ pc, while \form\ favored smaller \rinf. 
For factor of 3 changes, the fit degraded substantially (Fig. \ref{cscomp}). 
For $\rinf = 0.01$ pc, optically thick lines were too narrow
and the two peaks were nearly equal in strength, unlike the observations.
For $\rinf = 0.09$ pc, those lines were too wide and the blue/red ratio was too
large. There was also a greater conflict between the requirements of optically
thick and optically thin lines; if the abundance was increased to match the
latter, the former became too strong. Within the constraints on abundances
that we imposed, infall radii different by a factor of 3 would be strongly
ruled out. The mean absolute deviations over {\it all} species (\mean{AD})
for these different models are listed in Table \ref{vartab}.

The constraints on the infall radius from the molecular line observations
are inconsistent with those found by modeling the continuum emission
(Shirley et al. 2002). The predicted intensity profiles of the model
from Choi et al. (1995) were too flat to match the observations at
850 and 450 \micron\ (see Fig. 6 and Table 3 of Shirley et al 2002). 
To make an inside-out collapse model fit the data,
Shirley et al. had to use a very small infall radius, $r = 0.0048$ pc,
more than 6 times smaller than the infall radius that matches the line
profiles. Our modeling confirms that this small infall radius cannot match
the line profiles. This fundamental discrepancy between the models of the
dust and molecular line emission will be discussed further in \S \ref{disc}.

Harvey et al. (2001) found that an inside-out collapse model with
$\rinf \sim 0.03$ pc fit the extinction data well, but only if the
density was increased everywhere by a factor of about 5. We tried this
model; decreases in abundances by about an order of magnitude were required for
most species to bring line strengths back to near observed values.
We had to decrease the CS abundance within \rdep\ to match the data, but
this change is not unreasonable. The average deviation is somewhat
larger than the standard model, but not terrible.
Without constraints on abundances, it
is hard to rule out variations in the physical model of this magnitude.
However, the shapes of the CS lines were not reproduced well 
(top panel in Fig. \ref{cscomp}), with blue/red ratios clearly less than 
the observations.

Moving farther afield, one may consider other collapse models. In some
sense, the opposite extreme to the Shu (1977) model is the Larson-Penston
similarity solution (Larson 1969, Penston 1969). Line profiles from this
model were generated by Zhou (1992) and found to be considerably wider than
those observed in regions of low-mass star formation. More recently,
Masunaga et al. (1998) have shown that radiation hydrodynamical
(RHD) calculations of collapse are well approximated by a modified 
Larson-Penston model. 
This model produces lower infall velocities than the original Larson-Penston
model, which decrease with radius.
Masunaga \& Inutsuka (2000) have simulated line profiles from the RHD
models, finding blue profiles and smaller linewidths, qualitatively
consistent with those seen in low mass cores. While these models may
indeed have application in some regions, the linewidths listed in
Table 4 of their paper for models after formation of the central core
are larger than those in B335 by factors of at least two.

\begin{figure}
\plotfiddle{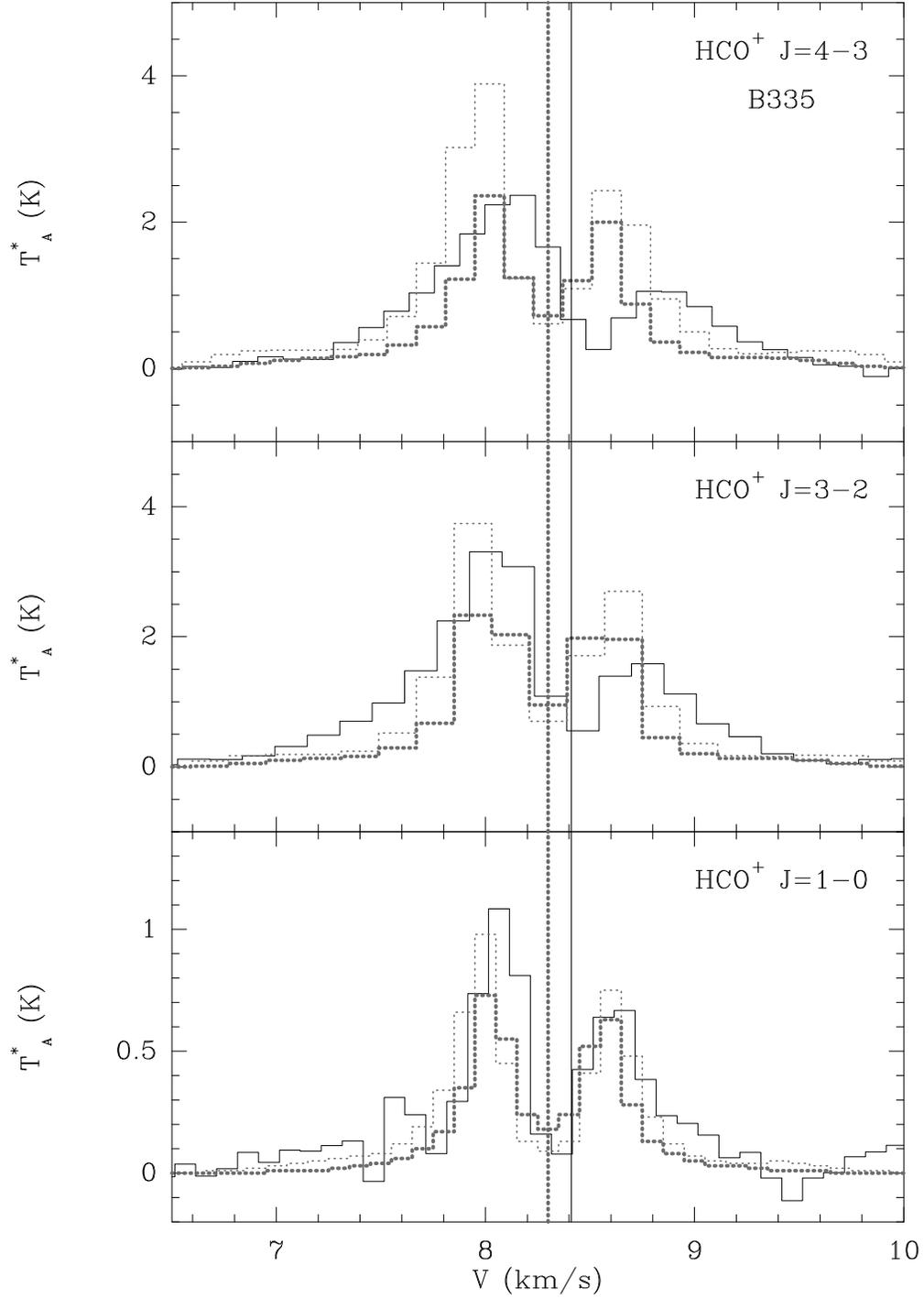}{7.0in}{0}{85}{85}{-250}{-65}
\caption{The observed \hcop\ lines are plotted on an expanded
scale, with the line profiles of the best step function model (dashed)
and chemical model (heavy dashed).
The solid vertical line is at 8.41 \kms, the average velocity of the dip,
while the dotted vertical line is at 8.30 \kms, the
mean velocity for optically thin lines. The shift of the dip velocity to
higher velocities seen in the observations is not reproduced in the models.
\label{vshift}
}
\end{figure}

There are hints in the spectra of deviations from the Shu model, particularly
in the shift of $v_{dip}$ to higher velocity for lines of higher excitation.
This shift can be see more clearly in Figure \ref{vshift}, where three lines
of \hcop\ are shown. The best step function model is also shown; it does not
reproduce this shift. The dip is caused by absorption from low-excitation
material. In the Shu model, the outer, static envelope is dominating this
absorption. A model with inward motion in this outer layer might better 
reproduce this shift.

\subsection{Self-consistent Chemical Models}

The chemical models are constrained by assuming an entire evolutionary
history for the core, as detailed by Lee et al. (2004). 
We consider only their standard model of the evolution of physical conditions
and luminosity to define the physical conditions, including the dust 
temperature profile, at the time step for which $\rinf=0.03$ pc.
Compared to the
standard model of Lee et al., we allowed adjustment of only 3 free
parameters: the binding energy to the dust, set by the assumed nature
of the dust surface; the initial abundance of elemental sulfur; and 
the cosmic-ray ionization rate.
The different models are summarized in Table \ref{chemmods}.

First, binding energies of molecules onto three different dust grain surfaces
were checked. For this comparison, the initial elemental abundances and 
the cosmic-ray ionization rate were the same as 
those in the standard model of Lee et al. (2004).
For the CS lines, the binding energy onto a CO-dominant grain mantle works the
best and the value of \mean{AD} is slightly better than for SiO$_2$. 
However, the low binding energy of molecules onto the CO mantle leads to
less freeze-out of CO, and, in turn, \nthp\ is destroyed by abundant CO in
the gas phase. As a result, simulated CO isotopologue lines are too strong, 
and simulated \nthp\ lines are too weak compared to the observations.
Attempts to improve the fit by reducing the initial abundance of carbon
make the fit to CS worse while still not making the models fit CO and \nthp\
profiles.

At the other extreme, CO and CS are frozen-out significantly onto 
\water -dominant grain mantles to produce much weaker lines than the 
observations indicate.
\hcop, which is a daughter molecule of CO, is also depleted from the gas phase.
Although \nthp\ is less likely to be destroyed by CO, even nitrogen 
molecules are easily frozen-out on the \water\ mantle, decreasing 
the \nthp\ abundance.
In addition, HCN increases by 3 orders of magnitude at radii smaller than 0.004
pc, compared to the abundance in the outer regions, giving a very broad line
wing, which is not present in the observed lines. 
Except for weaker CS lines, the lines simulated with abundance profiles 
calculated for bare SiO$_2$ grain surfaces show much 
better fits to actual data than do those from other assumptions about
grain surfaces.

We adopt the bare SiO$_2$ grain surface as our standard.
In this model, the CO evaporation radius is about 0.006 pc. At radii less 
than this radius, almost all CO is desorbed from dust grain surfaces.
Next, we varied the initial abundance of sulfur to improve the fit to the 
CS lines. 
An increase of the initial sulfur abundance by a factor of 5 gives the best
results with the SiO$_2$ grain surfaces. 
Other molecular lines do not vary much with the initial abundance of sulfur.
The abundance profiles of this model are shown as solid lines in 
Figure \ref{chem2fig}.
In all chemical models, the \nthp\ and \hcop\ lines are weaker than the 
observed lines, so we tested various cosmic-ray ionization rates.
A cosmic-ray ionization rate increased by a factor of 2 produced the best fit
for \nthp\ and \hcop\ lines.  We increased the ionization rate in the
energetics calculation for consistency.
HCN lines become stronger with the cosmic-ray ionization rate, and
\mean{AD} becomes somewhat worse if we 
increase the ionization rate by a factor of 5.
The abundances for the model with ionization enhanced by a factor of 2 
are shown by the dotted lines in Figure \ref{chem2fig}.

The chemical models produce abundances with large (many orders of magnitude)
variations with radius and quite complex radial structure. For explanations
of these effects, see Lee et al. (2004). Note in particular the large decreases
in abundance at large radii for most species caused by photo-dissociation.
The large decrease in CO abundance over a wide range of radii reflects
freeze-out onto grain surfaces, and some other species follow this trend,
but \nthp\ behaves oppositely because CO destroys \nthp. Likewise, most
species show a peak at small radii, where CO evaporates, because those
species also evaporate there, while \nthp\ decreases when CO evaporates.
The abundances for the step function model are also shown in Figure 
\ref{chem2fig}. In some cases, they are dramatically different. 

The line profiles resulting from the best fit chemical model are shown
in Figures \ref{csfig} to \ref{cnfig} as dotted lines. 
The average values of absolute deviation for the best chemical model
are also listed in Table \ref{abun} for comparison to those from the
step function models. In most cases, the values of $AD$ are slightly worse, most
notably for \dcop\ and \form, but many are similar to those of the step
function models.
Compared to the best fit with step function abundances, the best fit with 
chemical abundances shows less deep absorption dips in the CS 
lines, weaker lines of the higher transitions in \nthp\ and \form, and
less blue asymmetry in \hcop\ lines.   
In addition, the predicted lines of \hcop\ isotopologues are weaker than
in the step function models. To match the observations, the isotope ratios
would need to be increased by factors of 3 to 5. 
The deuterium ratio for the \dcop\ 3$-$2 would also need to be increased
by a factor of 5 to match the observed line. These discrepancies result
from the fact that B335 has quite strong lines of rare isotopologues of
\hcop.  However, standard isotope ratios for \cooo, \coo, and \coooo, 
or \form\ and \formi\ produce reasonably good matches to the observations.
Chemical abundances predict still weaker \form\ lines in high excitation 
transitions than do step function abundances.
As mentioned in \S \ref{step}, higher densities at small radii are necessary to 
account for the weak high excitation \form\ lines.  Also, the \nthp\ 
3$-$2 line is relatively weaker than the 1$-$0 line when compared to the 
observed lines, again suggesting the presence of higher densities at small 
radii. 
\hcop\ and HCN lines simulated with chemical abundances are narrower than
 the observed lines. 
These two molecules are abundant in outflowing gas.  
Therefore, the lines might be affected by the outflow, which is not considered
in this work. 
The satellite hyperfine lines of HCN 1$-$0 are produced better with the 
abundance profile from the chemical model than with a step function.
However, even chemical abundances cannot predict the satellite hyperfine
lines of the HCN 3$-$2 as well as the absence of the red component of the main
group. This result suggests the existence of a region with very high
HCN abundance.
The chemical models also do not reproduce the shift of $v_{dip}$ with
increasing $J$ seen in the \hcop\ (Fig. \ref{vshift}).

We also tested different time steps in the same luminosity model. In time
steps earlier than the time step for $\rinf=0.03$ pc, the model core has
higher densities at small radii, and the total internal luminosity is smaller 
than observed for B335 and vice versa for $\rinf > 0.03$ pc.
According to the test, \form\ and \nthp\ lines are fitted much better with
the chemical abundances in an earlier time step for $\rinf=0.015$ pc.
The satellite hyperfine lines of the HCN 3$-$2 are also well fitted, while 
its main group is too strong. Higher densities at small radii cause these 
results. In addition, at this time, the CO evaporation radius is about 
0.004 pc, so less CO evaporates compared to the time step for $\rinf=0.03$ pc.
As a result, \nthp\ is more abundant.
However, at this time, the infall velocities are not big enough to produce 
the degree of the blue asymmetry in CS lines. The CS lines predicted by
models with three times smaller and larger \rinf\ are also shown in
Figure \ref{cscomp} and the values of \mean{AD} are given in 
Table \ref{vartab}; unlike the step function models, the luminosity
is different for these other values of \rinf, because the luminosity
increases with time. Nonetheless, similar problems to those encountered
in the step function models appear at earlier and later times.

\section{Discussion} \label{disc} 

Both step function models and evolutionary chemical models do a reasonable
job of fitting most of the data. Models with constant abundances
are not adequate for fitting most observations.
In addition, the evolutionary chemical
models clearly indicate that abundances vary by orders of magnitude as
a result of freeze-out and desorption (Lee et al. 2004).
These conclusions are similar to those of J{\o}rgensen et al. (2004),
who find that constant abundance models are unsatisfactory and that a drop
function works better. The drop function, though simpler, is similar in shape
to the abundance profile of CO in Figure \ref{chem2fig}.
It allows a region of lower abundance at intermediate radii and a return
to high abundances at small radii. The drop function cannot of course
capture the full complexity of the abundance profiles in Figure \ref{chem2fig}.

Both models fit the CS lines despite the very different radial dependences
of the CS abundance. The \form\ lines are better fitted by the step function
models, primarily because they allow a high abundance over a substantial range
of radii, where the density is high, thus providing stronger lines of 
high-excitation transitions. Similarly, the chemical models, even with
enhanced ionization, cannot produce sufficient \hcop\ to match the
observations of rare isotopologues.
Interestingly, the chemical models do better on the HCN \jj10 line, but
neither model can match the peculiar \jj32 line profile.

In comparing models, one should bear in mind that the step function models
were allowed 15 free parameters, while the chemical models enjoyed only
3, and those were restricted by prior knowledge. In fact, the step function
abundances that fit best are very inconsistent with what we know of
chemistry. The CO abundance is depleted inside \rinf, while the \hcop\
and \form\ abundances increase; this combination is highly unlikely, especially
for \hcop, which is a direct product of CO. The very high abundance of
\hcop\ inside \rinf\ invoked by the step function models to match the 
strong lines of rare isotopologues is very hard to produce in any
chemical model.

The evolutionary model does not include grain surface reactions, so it does
not predict an ortho-para ratio for \form; we assumed a ratio of 1.7.
The most likely route for modifications to the chemical models
is to add grain surface reactions, but this step will effectively add
many free parameters to the chemical model because rates for surface
reactions are poorly known.

On the whole, it is in fact rather remarkable that the line profiles
from the self-consistent chemical models are as close as they are to
the observations. With variations in abundances of many orders of
magnitude with radius, they could easily have failed to match observations
by orders of magnitude. In addition, B335 is only one source, and it
has a rather rich spectrum for a low mass core, including unusually
strong lines of HC$^{18}$O$^+$. Of course, this feature makes B335 an attractive
source to test theories, but it also may mean that it is atypical.

What can explain the remaining differences between the models and the data?
First, the Shu (1977) model of the infall may not be correct. 
There is a hint in this direction in the fact that the models do not
reproduce the shift of $v_{dip}$ to the red of \vlsr\ for optically thin lines
(Fig. \ref{vshift}). This result suggests that the outer envelope is not
stationary, as in the Shu solution, but is also moving inward.
Second, the outflow is not considered in these models, but it will
affect abundances and line profiles. To include the outflow, we must
move beyond 1-D models.
Third, the chemical models may not yet include enough processes for
desorption of gas from grain mantles. These could increase the abundances
at intermediate radii.

\subsection{Off-Center Spectra}

We have focused here on spectra toward the center of B335. The large
differences between the empirical and evolutionary models in the abundances
of some species at larger radii suggest that spectra off the center of
B335 may help in testing chemical models. Figure \ref{off} shows a sample
of spectra at positions off the center predicted by the best-fit empirical and
evolutionary models. For \hcop\ and \cooo, we show observed spectra
at these positions, produced by averaging spectra displaced in all directions
with that separation in our maps. 
We also show predictions of CS spectra displaced by 60\arcsec;
we no longer have the observed spectra at those positions, but they are
presented by Zhou et al. (1993). For the \hcop\ \jj43 line, the evolutionary
model gives a weaker line, closer to the observations, than the empirical
model at 15\arcsec\ offset. Conversely, the evolutionary model produces
stronger lines of \cooo\ at 30\arcsec\ offset than either the empirical
model or the data. The differences are most dramatic for CS lines 60\arcsec\
away from the center. The strong decrease in CS abundance in the evolutionary
models, seen in Fig.  \ref{chem2fig}, produces lines that are much weaker 
than those of
the empirical model. Zhou et al. (1993) detected lines stronger than either
prediction, but the noise was fairly high. Improved maps of CS with
good spatial resolution would be very helpful in further constraining models.
The decrease in CS abundance at large radii is quite sensitive to the
external extinction (see Fig. 13 of Lee et al. 2004). Models with higher
values of external extinction, but with all other parameters unchanged,
do greatly increase the strength of CS lines at off-center positions.
Maps of appropriate
lines can help to constrain the environment of the core, while observations
with much higher resolution can test the predictions of abundance peaks
at small radii (Lee, Evans, \& Bergin 2005).

\subsection{What about the Dust?}

We noted above that the molecular line emission and the dust emission
lead to inconsistent conclusions about the density distribution. To
summarize, the dust emission data is consistent with an inside-out
collapse model only if the infall radius is much smaller than the molecular
line data indicate (Shirley et al. 2002). Attempts to adjust
abundances to make the line profiles predicted for the small infall radius
match the observations failed; the predicted line profiles are simply
too narrow at the early times implied by the small infall radius.

The other possibility is that the model based on the dust emission 
is wrong. While Shirley et al. performed an extensive set of tests,
there are two possibilities that remain to be considered: changes in the
dust opacity as a function of radius; and contributions to the
dust emission from a disk. Both of these would work by adding flux at
small radii, steepening the predicted radial intensity profile.
For example, Young et al. (2003) found that one could overestimate
$p$, the best-fit exponent in a power-law model by up to 0.5, if the
contribution of a disk was ignored. This difference could change the
best-fit value of $p = 1.8$ to something more compatible with the inner
part of an inside-out collapse.  

In fact, Harvey et al. (2004) have found evidence for a disk in B335 and
modeled multi-configuration data from the IRAM Plateau de Bure Interferometer
with an envelope-disk combination. They find a good fit with a disk
producing a flux of 21 mJy at 1.2 mm and an envelope with a broken power
law: 
$n = 3.3\ee4  (r/r_b)^{-1.5}$ for $r < r_b$; 
and 
$n = 3.3\ee4  (r/r_b)^{-2.0}$ for $r > r_b$. 
The best fit $r_b$ is 0.032 pc, essentially the same as our best fit
for the infall radius, and the two broken power laws agree with the
inside-out model for $r > \rinf$, and with the
asymptotic behavior of inside-out collapse at small radii.
However, the density just inside the infall radius has a slower dependence
on $r$ in the inside-out collapse solution; Harvey et al. (2004) note that
this theoretical $n(r)$ does not fit their data well. 

We have scaled up the flux of the disk in the Harvey et al. (2004)
model ($S_\nu \propto \nu^3$) and modeled the 850 and 450 \micron\
data from Shirley et al. (2000) with an inside-out collapse model
and a point source with the flux of the disk.
The intensity profiles are not well-fitted unless the point source is
5-10 times stronger than in the Harvey et al. model.
Models with disks and radial variations in dust opacity are needed to resolve
these questions, but they are outside the scope of this paper.

We conclude that the issues of disks (or more generally, compact structure)
and radial variations in dust opacity introduce enough uncertainty that
it is not yet possible to close the loop fully on the modeling of molecular
lines and dust continuum.

\section{Conclusions}

We have assembled data on a large number of molecular lines toward B335
and compared those lines to predictions of models. The models are of two
primary types: those with step function chemical abundances and those with
chemical abundances resulting from an evolutionary calculation (Lee et al.
2004). In both cases, the temperatures of dust and gas are calculated from 
the luminosity of the protostar and the density distribution under study.
In both cases, the underlying physical model is the inside-out collapse
model of Shu (1977).

Both step function and evolutionary chemical abundances can fit most of the
data, with some residual puzzles remaining. Both favor an infall radius
around 0.03 pc, as was found from earlier modeling by Zhou et al. (1993)
and Choi et al. (1995). Models with the same infall radius, but with densities
enhanced by a factor of 5, as suggested by Harvey et al. (2001) do not fit
as well, but they cannot be ruled out. Models with much smaller infall radii,
as favored by the dust continuum modeling (Shirley et al. 2002) do not
fit the data well at all. Resolving this discrepancy between the conclusions of
modeling dust and gas may require modifications to the dust models, including
incorporation of compact sources and changes in the dust opacities with
radius.

\begin{figure}[hbt!]
\plotfiddle{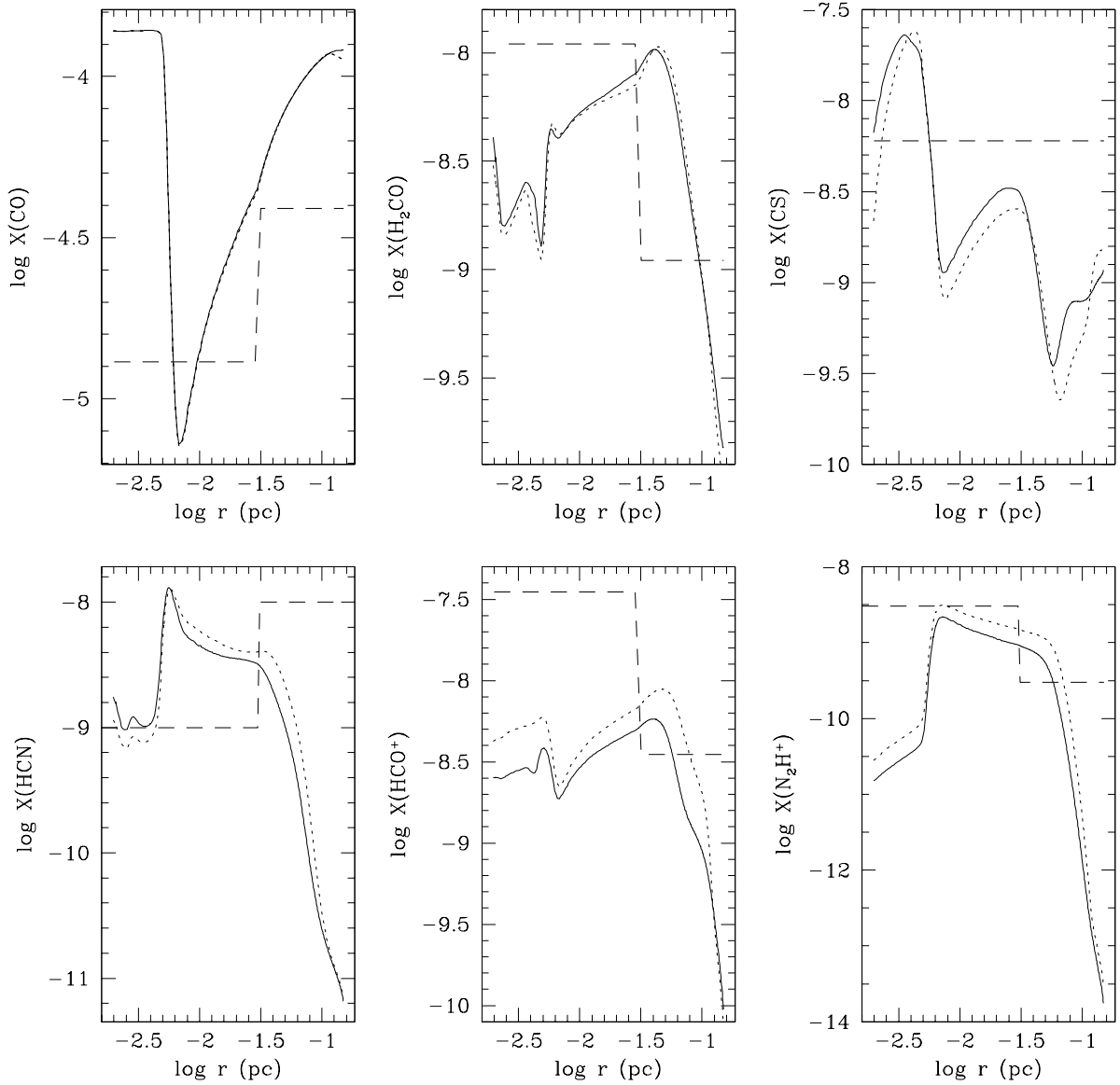}{7.0in}{0}{85}{85}{-250}{-75}
\caption{Abundance profiles. The dashed line shows the step function abundances,
and the solid and dotted lines represent the abundances calculated from the 
chemical evolution model of Lee et al. (2004) in the time step 
for $\rinf=0.03$ pc.
In both chemical models, we used the initial sulfur abundance 
greater than the standard value in Lee et al. (2004) by a factor of 
5. In the chemical model with dotted lines, the cosmic-ray ionization rate is
two times greater than the standard value to give the best fit to observed
line profiles (Model 6 in Table \ref{chemmods}).  The solid line is Model 5, 
which uses the standard value for ionization rate.
\label{chem2fig}
}
\end{figure}

\begin{figure}[hbt!]
\plotfiddle{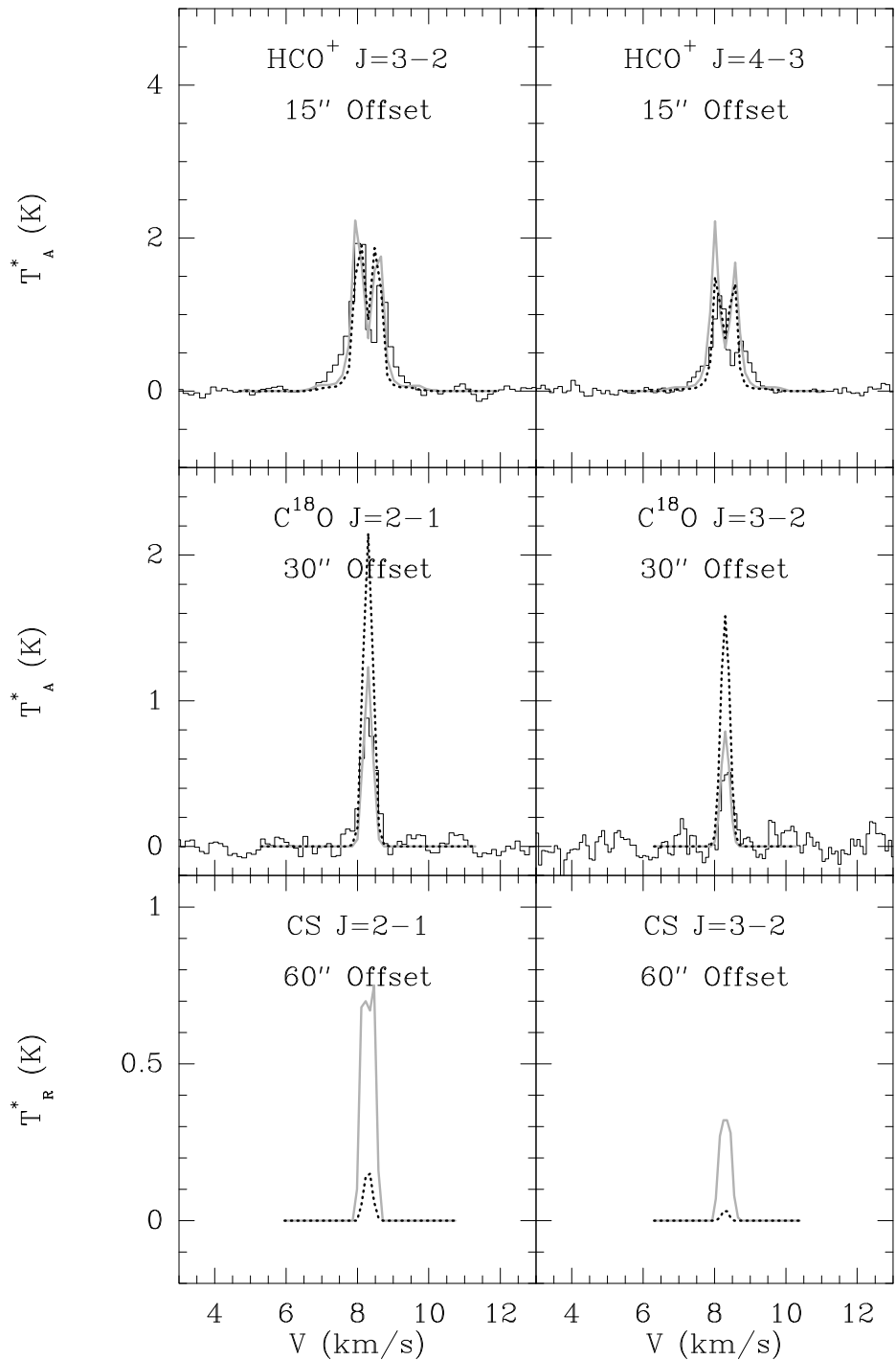}{7.0in}{0}{85}{85}{-250}{-65}
\caption{Off-center line profiles. 
The gray solid (step functions) and dotted (chemical
abundances) lines are predictions of the radiative transfer model.
For \hcop\ and \cooo, we show observed line profiles as histograms.
The observations of the CS lines can be seen in Zhou et al. (1993).
\label{off}
}
\end{figure}

The standard chemical evolution model shows huge variations in abundance
as a function of radius (Fig. \ref{chem2fig} and Lee et al. 2004), but
still comes reasonably close to matching the observations. This is quite
a remarkable fact. Changes to the initial sulfur abundance and cosmic ray
ionization rate improve the fit to the lines, but these may simply be
compensating for remaining unknowns in the chemistry. Rawlings and Yates
(2001) have highlighted the extreme sensitivity of some abundances and
line profiles to free parameters, especially the early evolutionary
history. Accordingly, the reader is cautioned that other combinations
of history, dynamics, and chemical parameters that we have not explored
might also produce reasonable fits to the data. The important point
is that chemical models now come close enough that one can begin to 
look in detail at what might improve the match. However, this should be
done after more than one source is compared to the models, as there will
be variations in conditions and evolutionary history from source to source.

In addition, the standard physical model of inside-out collapse does a
remarkably good job of predicting the line profiles. However, there are 
clear hints of dynamics beyond the standard model in the shift of the
velocity of the self-absorption dip to higher velocities in lines 
requiring higher excitation. Models with envelopes moving inward may
be more successful in reproducing these features. Spectra at positions
away from the center can constrain other parameters, especially the
surrounding radiation environment. However, future work
should also account for the non-sphericity and other effects of the outflow
in this source. Further observations with better spatial resolution will
be important to constrain these models, as the line profiles become
more diagnostic of both dynamics (Choi 2002) and chemistry (Lee et al.
2004).

\acknowledgments

We thank the referee for perceptive comments that led to improvements in
the paper.
We thank Y. Wang, who collected the Haystack data and who was involved in the
initial work on this project. We also thank Y-S. Park and W. Irvine for
supplying data from their papers. This research was supported in part by 
NSF grants AST-9988230 and AST-0307250 to the University of Texas at Austin.


\clearpage

\clearpage


\begin{table}[h]
\caption{Observing Parameters} \label{telescopes}
\vspace {3mm}
\begin{tabular}{l c c c c  c r r}
\tableline
\tableline
Line & $\nu$ & Telescope & $\eta_{mb}$ & $\theta_b$ &  $\delta v$ &Ref 
&Date \cr
& (GHz)  & &  & ($\arcsec$)  &  (km s$^{-1}$) & & \\ \tableline

CI $1\tto 0$     & 492.1607   & CSO  &   0.47 &   16  & 0.080 & 1 & 1996 Jun \cr
CN $2\tto 1$     & 226.874745 & CSO  &   0.60 &   27  &  0.14 & 1 & 1998 Jul \cr
C$^{17}$O $2\tto 1$ & 224.714368$^a$ & CSO & 0.81 &   33  &  0.17 & 1 & 2000 Jun \cr
C$^{18}$O $2\tto 1$ & 219.560352 & CSO & 0.57 &   28  &  0.15 & 1 & 1998 Jul \cr
C$^{18}$O $3\tto 2$ & 329.3305453 & CSO & 0.82 &  26  &  0.10 & 1 & 2000 Jul \cr
\hcop\ $1\tto 0$ &  89.188512 & Haystack & 0.12 & 25 & 0.10  & 1 & 1994 Jun \cr
\hcopi\ 1\tto0   &  86.754330 & Haystack & 0.12 & 25 & 0.10  & 1 & 1994 Jun \cr
HCO$^+$ 3\tto2      & 267.557619 & CSO   & 0.65 & 26  & 0.18  & 1  &1995 Mar \cr
H$^{13}$CO$^+$ 3\tto2 & 260.255339  & CSO  & 0.65 & 26 & 0.18 & 1 & 1995 Mar \cr
HC$^{18}$O$^+$ 3\tto2 & 255.47940   & CSO  & 0.65 & 26 & 0.18 & 1 & 1995 Mar \cr
HCO$^+$ 4\tto3   & 356.734288  & CSO    & 0.61 & 20 & 0.14    & 1 & 1995 Mar \cr
DCO$^+$ 3\tto2   & 216.112604  & CSO    & 0.57 & 28 & 0.16    & 1 & 1998 Jul \cr
HCN 1\tto0       & 89.635847   & TRAO   & 0.40 & 61 & 0.068   & 2 & 1997  \cr
HCN 3\tto2       & 265.8864343 & CSO    & 0.65 & 23 & 0.15    & 1 & 1996 Jun \cr
H$^{13}$CN 3\tto2 & 259.011814 & CSO   & 0.65 & 23 & 0.15    & 1 & 1996 Jun \cr
HNC 3\tto2       & 271.981142  & CSO    & 0.62 & 22 & 0.11    & 1 & 1996 Jun \cr
HNC 4\tto3       & 362.630303  & CSO    & 0.53 & 19 & 1.62    & 1 & 1997 Jul \cr
HC$_3$N 24\tto23 & 218.324788  & CSO    & 0.56 & 28 & 0.18    & 1 & 1996 Jun \cr
\nthp\ 1\tto0 & 93.176258$^b$ & Haystack & 0.12 & 25 & 0.10   & 1 & 1994 Jun \cr
\nthp\ 3\tto2 & 279.511757$^c$ & CSO    & 0.56 & 22 & 0.12    & 1 & 1996 Oct \cr
\ntdp\ 3\tto2 & 231.321775     & CSO    & 0.73 & 32 & 0.21    & 1 & 2001 Jul \cr
\pformi\ $1_{01}-0_{00}$ & 71.02478  & NRAO & 0.95 & 89 & 0.206  & 3 & \cr
\formi\ $2_{12}-1_{11}$ & 137.44996  & NRAO & 0.72 & 42 & 0.1065 & 3 & \cr
\form\ $2_{12}-1_{11}$  & 140.839518 & IRAM & 0.68 & 17 & 0.083  & 4 & \cr
\pformi\ $2_{02}-1_{01}$ & 141.98375 & NRAO & 0.72 & 42 & 0.103  & 3 & \cr
\formi\ $2_{11}-1_{10}$  & 146.63569 & NRAO & 0.72 & 42 & 0.0998 & 3 &  \cr
\form\ $3_{12}-2_{11}$  & 225.697772 & IRAM & 0.50 & 12 & 0.066  & 4 &  \cr
\form\ $3_{12}-2_{11}$  & 225.697787 & CSO  & 0.65 & 27 & 0.127  & 1 & 1996 Jun \cr
\pform\ $3_{03}-2_{02}$ & 218.222186 & CSO  & 0.65 & 28 & 0.131  & 1 & 1996 Jun \cr
\form\ $5_{15}-4_{14}$  & 351.768645 & CSO  & 0.53 & 20 & 0.083  & 1 & 1997 Jun \cr
\pform\ $5_{05}-4_{04}$ & 362.3530480 & CSO & 0.53 & 19 & 0.085  & 1 & 1997 Jun \cr
\pform\ $5_{23}-4_{22}$ & 365.3634280 & CSO & 0.53 & 19 & 0.085  & 1 & 1997 Jun \cr

\end{tabular}
\tablecomments{
(a) Reference frequency for the hyperfine shifts in Ladd et al. (1998);
(b) For the isolated hyperfine component (Lee et al. 2001); 
(c) Reference frequency for the hyperfine shifts in Caselli et al. (2002);
(1) This paper;
(2) Park et al. 1999;
(3) Minh et al. 1995;
(4) Zhou et al. 1993. 
}

\end{table}

\begin{table}[h]
\caption{Observational Results} \label{resultstab}
\vspace {3mm}
\begin{tabular}{l l r r r r }
\tableline
\tableline

Molecule  & Line     & \intint     & \ta       & $v_{LSR}$ &  $\Delta v$ \cr
          &          & (K \kms)    & (K)       & (\kms)    &  (\kms)      
\\ 
\tableline
CI        &\jj10     & 0.80(0.04)  & 1.51(0.18)& 8.29(0.01)& 0.50(0.03)   \cr
CN$^{b,c}$ &\jj21    & 0.90(0.02)  & 0.43(0.02)& 8.35(0.04)& 0.62(0.07)   \cr
C$^{17}$O$^a$ &\jj21 & 0.46(0.07)  & 0.57(0.07)& 8.39(0.02)& 0.49(0.07)   \cr
C$^{18}$O &\jj21     & 0.63(0.42)  & 1.10(0.10)& 8.27(0.02)& 0.57(0.04)   \cr
C$^{18}$O &\jj32     & 0.78(0.09)  & 2.15(0.26)& 8.33(0.04)& 0.64(0.08)   \cr
\hcop$^b$ &\jj10     & 0.71(0.05)  & 1.15(0.11)& 8.35(0.03)& 0.61(0.10)   \cr
\hcopi    &\jj10     & 0.24(0.06)  & 0.33(0.02)& 8.23(0.04)& 0.58(0.03)   \cr
\hcop$^b$ &\jj32     & 3.13(0.05)  & 3.32(0.08)& 8.46(0.04)& 0.94(0.16)   \cr
\hcopi    &\jj32     & 0.46(0.01)  & 0.76(0.03)& 8.25(0.01)& 0.57(0.02)   \cr
\hcopii   &\jj32     & 0.05(0.01)  & 0.09(0.01)& 8.26(0.04)& 0.58(0.10)   \cr
\hcop$^b$ &\jj43     & 2.27(0.15)  & 2.35(0.06)& 8.55(0.03)& 0.97(0.12)   \cr
\dcop     &\jj32     & 0.78(0.03)  & 1.32(0.04)& 8.35(0.01)& 0.55(0.02)   \cr
HCN$^{b,c}$ &\jj10     & 0.59(0.10)  & 0.35(0.05)& 8.39(0.01)& 0.70(0.10)   \cr
HCN$^b$   &\jj32     & 0.81(0.03)  & 0.80(0.02)& 8.40(0.20)& 1.01(0.04)   \cr
H$^{13}$CN&\jj32     & \nodata     & $<0.1$    & \nodata   & \nodata      \cr
HNC       &\jj32     & 0.22(0.01)  & 0.49(0.03)& 8.16(0.01)& 0.42(0.03)   \cr
HNC       &\jj43     & 0.39(0.05)  & 0.16(0.03)& 8.33(0.16)& 2.3(0.4)     \cr
HC$_3$N &\jj{24}{23} & \nodata     & $<0.08$   & \nodata   & \nodata      \cr
\nthp$^d$ &\jj10     & 0.13(0.01)  & 0.25(0.02)& 8.33(0.01)& 0.47(0.02)   \cr
\nthp$^{b,c}$ &\jj32  & 0.95(0.08) & 1.00(0.04)& 8.38(0.03)& 0.38(0.04)   \cr
\ntdp$^a$ &\jj32      & 0.32(0.04) & 0.38(0.04)& 8.36(0.02)& 0.31(0.05)   \cr
\form  &\jkkjkk312211 & 0.55(0.01) & 0.63(0.02)& 8.26(0.01)& 0.82(0.02)   \cr
\pform &\jkkjkk303202 & 0.45(0.01) & 0.59(0.03)& 8.28(0.01)& 0.71(0.03)   \cr
\form  &\jkkjkk515414 & 0.49(0.03) & 0.57(0.07)& 8.30(0.02)& 0.80(0.07)   \cr
\pform &\jkkjkk505404 & 0.40(0.06) & 0.31(0.11)& 8.53(0.08)& 1.22(0.23)  \cr
\pform &\jkkjkk523422 & \nodata    & $<0.2$    & \nodata   & \nodata     \cr

\end{tabular}
\tablecomments{(a) Hyperfine structure: \intint\ refers to area under all lines;
\ta\ refers to main peak of blended components; $v_{LSR}$ and $\Delta v$
are from fit to blended components;
(b) Double-peaked: \intint\ is for total area under both peaks;
\ta\ refers to strongest peak; $v_{LSR}$ refers to dip;
$\Delta v$ is integrated intensity divided by peak \ta.
(c) Hyperfine structure: \intint\ refers to area under all lines;
$\Delta v$ determined from isolated hyperfine component.
(d) All entries refer to the isolated hyperfine component.
}

\end{table}


\begin{table}[h]
\caption{Standard Physical Model } \label{physmod}
\vspace {3mm}
\begin{tabular}{c c c c c c c c c c c}
\tableline
\tableline
 Type &$a$     & \rinf & \rout &$L$       & \av & \go  & $\zeta$ &$b$& \nel & $X(CO)$   \cr
    & (\kms) &  (pc) &  (pc) & (\lsun)  & (mag) &   & (s$^{-1}$) & (\kms)& (\cmv) &    \\ 
\tableline
 Shu   & 0.23   & 0.03  & 0.15  & 4.5      & 1.3  & 0.1 & 3\ee{-17} & 0.12 & 1\ee{-3} & 7.4\ee{-5}    \cr
\tableline
\end{tabular}
\end{table}

\begin{table}[h]
\caption{Isotope Ratios } \label{isotopes}
\vspace {3mm}
\begin{tabular}{r r r   }
\tableline
\tableline
 C/$^{13}$C & O/$^{18}$O & $^{18}$O/$^{17}$O    \cr
\tableline
70            & 540  & 3.5  \cr
\tableline
\end{tabular}
\end{table}

\begin{table}[h]
\caption{Aggregated Hyperfine Shifts\tablenotemark{a} and 
Strengths\tablenotemark{b} } \label{hpftab}
\vspace {3mm}
\begin{tabular}{r r r r r r r r r r }
\tableline
\tableline
 C$^{17}$O &2\tto1 & HCN & 1\tto0 & HCN & 3\tto2 & \nthp & 1\tto0 
&\nthp &3\tto2    \cr
 \dv & $r_i$ & \dv & $r_i$ & \dv & $r_i$ & \dv & $r_i$ & \dv & $r_i$ \\
\tableline
1.157 & 0.040 & 4.849 & 0.333  & 1.749 & 0.037 & 6.936 & 0.037 &2.015 & 0.017   \cr
0.431 & 0.122 & 0.000 & 0.556  & 0.303 & 0.200  & 5.984 & 0.185 &0.669 & 0.015   \cr
0.241 & 0.571 & $-$7.072 & 0.111 & $-$0.030 & 0.725 & 5.545 & 0.111 &0.416 &0.084   \cr
$-$0.526 & 0.093 &\nodata &\nodata &$-$0.611 & 0.001 & 0.956 & 0.185 &0.266 & 0.094  \cr
$-$0.926 & 0.016 &\nodata &\nodata &$-$2.348 & 0.037 & 0.000 & 0.259 &0.076 & 0.089 \cr
$-$1.073 & 0.095 &\nodata &\nodata &\nodata &\nodata & $-$0.611 & 0.111 &$-$0.073 & 0.615 \cr
$-$1.203 & 0.062 &\nodata &\nodata &\nodata &\nodata & $-$8.006 & 0.111 & $-$0.601 & 0.010   \cr
\nodata &\nodata &\nodata &\nodata &\nodata &\nodata &\nodata &\nodata &$-$2.644 & 0.011  \cr
\nodata &\nodata &\nodata &\nodata &\nodata &\nodata &\nodata &\nodata &$-$2.773 & 0.014  \cr
\end{tabular}
\tablecomments{(a) Shift in \kms\ relative to the assumed central velocity
of 8.30 \kms;
(b) Relative strength ($r_i$) normalized so that $\sum r_i = 1$.
}
\end{table}

\begin{table}[h]
\caption{Step Function Abundances} \label{abun}
\vspace {3mm}
\begin{tabular}{l r r r r  }
\tableline
\tableline
Species &  $X(r>r_{inf})$ & $X(r<r_{inf})$ & AD(Step) & AD(Chem)    \cr
\tableline
CS            &6.0\ee{-9} & 6.0\ee{-9} & 6.88	& 6.78  \cr
\cooo\        &7.4\ee{-8} & 2.5\ee{-8} & 3.00	& 3.22  \cr
\hcop\        &3.5\ee{-9} & 3.5\ee{-8} & 3.95	& 4.02  \cr
\dcop\        &6.0\ee{-11} &6.0\ee{-10}& 1.87	& 3.27  \cr
\nthp\        &3.0\ee{-10} & 3.0\ee{-9}& 5.15	& 6.14  \cr
HCN           &1.0\ee{-8}  & 1.0\ee{-9}& 5.43	& 4.73  \cr
\form\        &7.0\ee{-10} & 7.0\ee{-9}& 2.64	& 3.22	\cr
\pform\       &4.0\ee{-10} & 4.0\ee{-9}& ...    & ...   \cr 
\tableline
\end{tabular}
\end{table}
\begin{table}[h]
\caption{Variations in the Physical Model} \label{vartab}
\vspace {3mm}
\begin{tabular}{r c r r r  }
\tableline
\tableline
 \rinf & Density factor & \mean{AD}\tablenotemark{a} & Chem Mod. & 
\mean{AD}\tablenotemark{a}   \cr
\tableline
0.01    & 1.0  & 4.62  & 6	& 4.93 \cr
0.03	& 1.0	& 3.75 & 6	& 4.15 \cr
0.09	& 1.0	& 5.38 & 6	& 4.95 \cr
0.03	& 5.0	& 4.02 & \nodata & \nodata \cr
\end{tabular}
\tablenotetext{a}{The mean value of AD for all species.}
\end{table}

\clearpage

\begin{table}[h]
\caption{Chemical Models } \label{chemmods}
\vspace {3mm}
\begin{tabular}{l r r r r r }
\tableline
\tableline
Model & Type  & Dust    & $X$(S) & $\zeta$ & \mean{AD}\tablenotemark{a}  \cr
      &       & Surface &        & (s$^{-1}$) &          \\ 
\tableline
Model 1  & Step  & \nodata & \nodata & \nodata & 3.75 \cr
Model 2  & Chem  & SiO$_2$ & 4\ee{-8}& 3.0\ee{-17}  & 5.05 \cr
Model 3  & Chem  & CO      & 4\ee{-8}& 3.0\ee{-17}  & 4.84 \cr
Model 4  & Chem  & H$_2$O  & 4\ee{-8}& 3.0\ee{-17}  & 5.89 \cr
Model 5  & Chem  & SiO$_2$ & 2\ee{-7}& 3.0\ee{-17}  & 4.25 \cr
Model 6  & Chem  & SiO$_2$ & 2\ee{-7}& 6.0\ee{-17}  & 4.15 \cr
Model 7  & Chem  & SiO$_2$ & 2\ee{-7}& 1.5\ee{-16}  & 4.23 \cr
\end{tabular}
\tablenotetext{a}{The mean value of AD for all species.}
\end{table}

\end{document}